\documentclass[preprint,showpacs,preprintnumbers,amsmath,amssymb]{revtex4-2}
\usepackage{graphicx}
\usepackage{epsfig}
\usepackage{bm}
\usepackage{amsfonts}
\usepackage{subfigure}
\usepackage{multirow}
\usepackage{float}
\usepackage{color}
\usepackage{hyperref}

\begin{document}

\title{\textbf{Primordial black holes in nonminimal derivative coupling inflation with quartic potential and reheating consideration}}


\author{Soma Heydari\footnote{s.heydari@uok.ac.ir} and Kayoomars Karami\footnote{kkarami@uok.ac.ir}}
\address{\small{Department of Physics, University of Kurdistan, Pasdaran Street, P.O. Box 66177-15175, Sanandaj, Iran}}
\date{\today}

\begin{abstract}

We investigate the generation of Primordial Black Holes (PBHs) with the aid of gravitationally increased friction mechanism originated from the NonMinimal field Derivative Coupling (NMDC) to gravity  framework, with the quartic potential. Applying the coupling parameter as a two-parted function of inflaton field and fine-tuning of five parameter assortments we can acquire ultra slow-roll phase to slow down the inflaton field due to high friction. This enables us to achieve enough enhancement in the amplitude of curvature perturbations power spectra to generate PBHs with  different masses. The reheating stage is considered to obtain criteria for PBHs generation during radiation dominated era. We demonstrate  that three cases of asteroid mass PBHs ($10^{-12}M_{\odot}$, $10^{-13}M_{\odot}$, and $10^{-15}M_{\odot})$ can be very interesting candidates for comprising $100\%$ , $98.3\%$ and $99.1\%$ of the total Dark Matter (DM) content of the universe. Moreover, we analyse the production of induced Gravitational Waves (GWs), and illustrate that their spectra of current density parameter $(\Omega_{\rm GW_0})$ for all parameter Cases foretold by our model have climaxes which cut the sensitivity curves of GWs detectors, ergo the veracity of our outcomes can be tested in light of these detectors.  At last, our numerical results exhibit that  the spectra of $\Omega_{\rm GW_0}$  behave as a power-law function with respect to frequency, $\Omega_{\rm GW_0} (f) \sim (f/f_c)^{n} $, in the vicinity of climaxes. Also, in the infrared regime $f\ll f_{c}$, the power index satisfies the relation $n=3-2/\ln(f_c/f)$.
 \end{abstract}
\maketitle

\newpage
\section{Introduction}
The concept of Primordial Black Holes (PBHs) generation  from the   over-dense regions of the early universe
should be addressed in the first instance to the recommendation by Zel'dovich and Novikov in 1967 \cite{zeldovich:1967}, and subsequent progressive researches by  Hawking and Carr in 1970s \cite{Hawking:1971,Carr:1974,carr:1975}.
Latterly, the prosperous discovery of  Gravitational Waves (GWs) arises from two coalescing black holes  with masses around  $ 30 M_\odot$ ($M_\odot$ denotes the solar mass) by LIGO-Virgo Collaboration \cite{Abbott:2016-a,Abbott:2016-b,Abbott:2017-a,Abbott:2017-b,Abbott:2017-c}, has attracted wide attention to PBHs as a possible  source of  the universe Dark Matter (DM) content and GWs \cite{Bird:2016,Clesse:2017,Sasaki:2016,Carr:2016,Belotsky:2014,Belotsky:2019}.
Inasmuch as the PBHs generate from the gravitationally collapse of the primordial density perturbations, hence they are not bounded by Chandrasekhar mass limitation owing to their non-stellar origin,  and have wide range of masses. In mass scale around ${\cal O}(10^{-5})M_\odot$,  PBHs  locate in the permitted region by OGLE data \cite{OGLE}.   Hence they  can be considered as the origin of ultrashort-timescale microlensing events in the OGLE data with the fractional abundance about ${\cal O}(10^{-2})$  of the total DM content.
By reason of the invalidity of gravitational femtolensing of gamma-ray bursts \cite{Barnacka} through the declined lensing  effects by the wave effects \cite{HSC,Katz},  the  Subaru Hyper Supreme-Cam (Subaru HSC) microlensing observations impose no restriction on PBHs with masses  less than $10^{-11}M_{\odot}$. Thereupon, considering  ineffectiveness of the restriction imposed by white dwarfs  \cite{WD} on  mass range $(10^{-14}-10^{-13})M_\odot$ due to numerical simulations in \cite{Montero-Camacho,Carr-Kuhnel}, the existence of PBHs in  the mass window  of asteroid masses  $(10^{-16}-10^{-11})M_\odot$ as interesting candidates  comprising all DM content of universe is possible \cite{Laha:2019}.

Generally, producing a large enough amplitude of primordial curvature perturbations (${\cal R}$) during inflationary epoch is necessary to generate PBHs during Radiation Dominated (RD) era. Overdense regions can be formed when superhorizon scales associated with the large amplitude of ${\cal R}$ become subhorizon during RD era, and gravitationally collapse of these overdensities generate PBHs.  Notwithstanding the restriction of the  power spectrum of ${\cal R}$ at large scales by CMB anisotropies normalization to $2.1 \times 10^{-9}$ \cite{akrami:2018},
PBHs generation requires an enhancement in the power spectrum of ${\cal R}$ to order  ${\cal O}(10^{-2})$ at scales smaller than CMB scales. Newly, different techniques for  multiplying the  amplitude of the power spectrum of ${\cal R}$ at small scale by seven orders of magnitude in comparison with CMB scales have been inspected in many investigation \cite{Khlopov:2010,Cai:2018,Ballesteros:2020a,Kamenshchik:2019,fu:2019,Dalianis:2019,mahbub:2020,mishra:2020,
Fumagalli:2020a,Liu:2020,Dalianis:2020,Heydari:2021,Solbi-a:2021,Solbi-b:2021,Teimoori-b:2021,Rezazadeh:2021,
Bhaumik:2020,Yogesh:2021,Domenech-a:2020,Domenech-b:2021,Kimura:2021,Kawai:2021,Kawai:2021edk,Lin:2020,Lin-b:2021,
Lu:2020,Zhang:2021,Domenech:2020b,fu:2020}. One of the proper ways to achieve a rise in the scalar  power spectrum is a  brief period of Ultra Slow-Roll (USR) inflation due to  declining the speed of inflaton field via gravitationally enhanced friction. The framework of NonMinimal Derivative Coupling to gravity (NMDC) beside the fine-tuning of the parameters of the model can give rise to increase friction gravitationally \cite{Germeni:2010,Defelice:2011,Tsujikawa:2012,Tsujikawa:2013,Defelice:2013,Dalianis:2020,fu:2020,fu:2019,Teimoori:2021,Heydari:2021}.

The nonminimal derivative coupling model is a subclass of a generic  scalar-tensor theory with second-order equations of motion namely Horndeski theory \cite{Defelice:2011,Tsujikawa:2012,Tsujikawa:2013,Defelice:2013,Horndeski:1974}, which prevents the model from negative energy and pertinent instabilities \cite{ostrogradski:1850}.  A characteristic  of the nonminimal field  derivative coupling to gravity  is that the gravitationally increased friction mechanism can be applied for generic steep potentials such as quartic potential \cite{Tsujikawa:2012}. In view of the Planck  2018  TT,TE,EE+lowE+lensing+BK14+BAO data the quartic potential in the standard inflationary model is incompatible with the  latest CMB data and completely  ruled out \cite{akrami:2018}. As regards a lot of effort have been put into making the Higgs field consistent with inflation,  One way  is employing the nonminimal field  derivative coupling to gravity \cite{Germeni:2010} (see  \cite{Amendola:1993} for original study). With this in mind, we contemplate the quartic potential in the nonminimal derivative coupling setup and try to rectify its observational results at CMB scales and examine the feasibility of PBHs generation of the mentioned mass scales simultaneously.

It is worth noting that an analogous investigation  has been done by Fu et al. \cite{fu:2019} and Dalianis et al. \cite{Dalianis:2020}.  In \cite{fu:2019} for the chosen parameter  the influence  of the temporary  nonminimal derivative coupling framework far from the peak position $(\phi=\phi_{c})$ is faded away and the standard slow-roll inflation is recovered. In the other word the NMDC framework just governs during USR phase. But in our work  the authentication of the  nonminimal derivative coupling framework throughout the inflationary epoch is respected because of the form of our suggested coupling function between field derivative and Einstein tensor.\\
Akin to \cite{Dalianis:2020} we choose quartic potential in  NMDC framework with the same coupling function, but in our model inflaton is not Higgs and formation of multifarious PBHs in asteroid, earth,  and stellar mass ranges thereto their concurrent scalar gravitational waves  are predicted. Moreover reheating consideration for our model terminate in obtaining  criteria for PBHs generation during RD era.

Generating  of  induced gravitational waves can be  another effect of  re-entry the large enough  densities of primordial perturbations to the horizon during the RD epoch, concurrent with the PBHs formation
\cite{Kohri:2018,Cai:2019-a,Cai:2019-b,Bartolo:2019-a,Bartolo:2019-b,Wang:2019,Cai:2019-c,Xu:2020,Lu:2019,
Hajkarim:2019,Fumagalli:2020b,Bhaumik:2021}.
Heretofore miscellaneous mechanisms  of  PBHs and detectable GWs generation owing to intensifying  the small scale primordial perturbations  have been suggested in many inflationary models.
The popular one of them is the  single-field  model with a potential having the flat region pertinent to USR phase in the environs of an inflection-point which  leads to a rise in the curvature perturbations and generate PBHs and GWs \cite{Motohashi:2017,Germani:2017,Di:2018}. Another models with Gaussian form of scalar power spectrum \cite{Namba:2015,Garcia-Bellido:2017, Lu:2019}, or multifield inflationary models such as double \cite{Kawasaki:2016,Kannike:2017}, and hybrid  \cite{Garcia-Bellido:1996,Clesse:2015} inflation  have been able to explicate  the formation of PBHs and induced GWs via enhancement of scalar power spectrum during the inflationary era.  In the following of this work, we  evaluate the current fractional density parameter of induced GWs in our NMDC setup and check the rectitude of  our numerical conclusion  with the latest  sensitivity curves of  GWs detectors.

The reminder  of this paper is organized as  follows. In Sec. \ref{sec2}, we study the basis formulae of the  nonminimal derivative coupling framework briefly. In Sec. \ref{sec3}, we explain the suitable mechanism of gravitationally increased friction to enhance the amplitude of the scalar power spectrum to around order ${\cal O}(10^{-2})$ in our model. To specify when the superhorizon climax scales of perturbations re-enter the horizon, we take into account reheating consideration in Sec. \ref{sec4}. In the succeeding sections we calculate different masses and fractional abundances for PBHs in Sec. \ref{sec5} and current energy spectra for induced GWs in Sec. \ref{sec6}. Ultimately, we epitomize our outcomes in Sec. \ref{sec7}.
\section{Foundation of Nonminimal Derivative Coupling framework}\label{sec2}
We initiate this section with the general action of the NonMinimal Derivative Coupling (NMDC) model, in which the inflaton field is nonminimally coupled to the Einstein tensor via its derivative as follows \cite{Germeni:2010,Defelice:2011,Tsujikawa:2012,Tsujikawa:2013,Defelice:2013}
\begin{equation}\label{action}
S=  \int {\rm d}^{4}x\sqrt{-g}\bigg[\frac{1}{2}R-\frac{1}{2}\big(g^{\mu\nu}-\xi G^{\mu\nu}\big)\partial_{\mu}\phi\partial_{\nu}\phi-V(\phi)\bigg],
\end{equation}
wherein $g$ is the determinant of the metric $ g_{{\mu}{\nu}}$, $R$ is the Ricci scalar, $G^{{\mu}{\nu}}$ is the Einstein tensor, $\xi$ is the coupling parameter with dimension of  $({\rm  mass})^{-2}$, and $V(\phi)$ is the potential of a scalar field $\phi$. As we explained in the previous section the action (\ref{action}) appertains to the Horndeski theory, which produces second order equations.
The Lagrangian of this class of theories consists of the term $G_{5}(\phi,X)G^{\mu\nu}(\nabla_{\mu}\nabla_{\nu}\phi)$, where $G_{5}$ is a general function of $\phi$ and kinetic term $X=-\frac{1}{2}g^{\mu\nu}\partial_{\mu}\phi \partial_{\nu}\phi$. By taking $G_{5}=-\Theta(\phi)/2$, then integrating by parts, and defining  $\xi\equiv d\Theta/d\phi$, the mentioned Horndeski's Lagrangian leads to the NMDC action  (\ref{action}). In \cite{Germeni:2010,Defelice:2011,Tsujikawa:2012,Tsujikawa:2013,Defelice:2013} the Case of constant $\xi$ has been investigated. Moreover, in \cite{Granda:2020} another investigation has been done by considering $\xi=\beta/\phi^{n}$, where $\beta$ is a parameter with dimension of $({\rm mass})^{n-2}$ and $n$ is an integer, in order to adapt the observational results of steep potentials like Higgs with the present CMB data.

In this work, we consider a more general and appropriate function of $\phi$ such as $\xi=\theta(\phi)$, so that not only the observational results of the quartic potential on large scales is reanimated, but also a  tenable outline for PBHs and induced GWs production on small scales can be achieved. For this purpose, we start to study the background evolution of the homogeneous and isotropic universe with the
flat Friedmann-Robertson-Walker (FRW) metric and mostly positive signature $g_{{\mu}{\nu}}={\rm diag}\Big(-1, a^{2}(t), a^{2}(t), a^{2}(t)\Big)$,
wherein  $a(t)$ is the scale factor and $t$ is the cosmic time. We also adhere to convention of reduced Planck mass is identical to unity $(M_P=1/\sqrt{8\pi G}=1)$, over all this paper.
In addition, we review the fundamental relations for produced perturbations during inflation epoch in our model described by action (\ref{action}).

The Friedmann equations and the equation of motion for $\phi$ can be obtained via taking variation of action (\ref{action}) with respect to $g_{\mu\nu}$ and $\phi$ as follows
\begin{align}
& 3H^{2}-\frac{1}{2}\Big(1+9H^{2}\theta(\phi)\Big)\dot{\phi}^{2}-V(\phi)=0,
\label{FR1:eq}
\\
& 2\dot{H}+\left(-\theta(\phi)\dot{H}+3\theta(\phi)H^{2}+1\right)\dot{\phi}^{2}-H\theta_{,\phi}\dot{\phi}^{3}
-2H\theta(\phi)\dot{\phi}\ddot{\phi}=0,
\label{FR2:eq}
\\
& \left(1+3\theta(\phi)H^{2}\right)\ddot{\phi}+\Big(1+\theta(\phi)(2\dot{H}+3H^{2})\Big)3H\dot{\phi}
 +\frac{3}{2}\theta_{,\phi}H^{2}\dot{\phi}^{2}+V_{,\phi}=0,
\label{Field:eq}
\end{align}
where $H\equiv \dot{a}/a $ is the Hubble parameter,  the dot indicates derivative with respect to the cosmic time $t$, and moreover the notation $({,\phi})$ indicates ${d }/{d\phi}$.
In the NMDC setup, the slow-roll parameters are specified as \cite{Defelice:2011,Tsujikawa:2012}
\begin{equation}\label{SRP}
  \varepsilon \equiv -\frac{\dot H}{H^2}, \hspace{.5cm}  \delta_{\phi}\equiv \frac{\ddot{\phi}}{ H\, \dot{\phi}}, \hspace{.5cm}\delta_{X}\equiv \frac{\dot{\phi}^2}{2 H^2}, \hspace{.5cm} \delta_{D}\equiv \frac{\theta(\phi)\dot{\phi}^2}{4}.
\end{equation}
It is known that, the validity of the slow-roll inflation is described by $\{\epsilon, |\delta_{\phi}|, \delta_{X},\delta_{D}\}\ll 1$. Under the slow-roll approximation, the energy density of the inflaton is dominated by the potential energy and the equations
(\ref{FR1:eq}), (\ref{FR2:eq}) and (\ref{Field:eq}) abbreviate to
\begin{align}
\label{FR1:SR}
& 3 H^2\simeq V(\phi),\\
  \label{FR2:SR}
& 2\dot{H}+{\cal A}\dot{\phi}^2-H \theta_{,\phi}\dot{\phi}^3\simeq0,\\
  \label{Field:SR}
& 3 H\dot{\phi}{\cal A}+\frac{3}{2}\theta_{,\phi}H^2\dot{\phi}^2+V_{,\phi}\simeq0,
\end{align}
wherein
\begin{equation}\label{A}
{\cal A}\equiv 1+3 \theta(\phi) H^2.
\end{equation}
Also because of the ease of calculation, we take the below assumption throughout the slow-roll domain
\begin{equation}\label{condition}
|\theta_{,\phi}H\dot{\phi}|\ll {\cal A}.
\end{equation}
By utilizing the assumption (\ref{condition}), the equations (\ref{FR1:SR}), (\ref{FR2:SR}) and (\ref{Field:SR}) take the following simpler form as
\vspace{-0.5cm}
\begin{align}
\label{FR1:SRC}
& 3 H^2\simeq V(\phi),\\
  \label{FR2:SRC}
& 2\dot{H}+{\cal A}\dot{\phi}^2\simeq0,\\
  \label{Field:SRC}
& 3 H\dot{\phi}{\cal A}+V_{,\phi}\simeq0.
\end{align}
Applying Eqs. (\ref{FR1:SRC}) and (\ref{Field:SRC}), we obtain
\begin{equation}\label{epsilon}
\varepsilon \simeq \delta_{X}+6\delta_{D}\simeq \frac{\varepsilon_{V}}{{\cal A}},
\end{equation}
where
\vspace{-0.5cm}
\begin{equation}\label{epsilonv}
\varepsilon_{V}\equiv \frac{1}{2}\left(\frac{V_{,\phi}}{V}\right)^2.
\end{equation}
Note that  $\varepsilon\simeq \varepsilon_{V}$ is concluded for ${\cal A}\simeq 1$, and one can attain  the restoration of the standard slow-roll
 inflationary formalism. As we can see from  (\ref{epsilon}), ${\cal A}\gg 1$ (high friction domain) results in  $\varepsilon \ll \varepsilon_{V}$, therefore  inflaton embogs owing to gravitationally increased friction, and rolls more slowly in comparison to that in the standard slow-roll story line. On the other hand, we expect that the severe   decline of $\varepsilon$ in the mentioned high friction domain can give rise to an increase in the amplitude of curvature perturbations $({\cal R})$. With this aspect in mind, we should inspect the power spectrum of the curvature perturbations, which is  calculated  by \cite{Tsujikawa:2013} at the moment of the Hubble radius crossing by the comoving wavenumber
($c_{s}k=a H$)  as follows
\begin{equation}\label{Ps}
{\cal P}_{s}=\frac{H^2}{8 \pi ^{2}Q_{s}c_{s}^3}\Big|_{c_{s}k=aH}\,,
\end{equation}
wherein
\begin{align}\label{Qs,cs2}
& Q_{s}= \frac{ w_1(4 w_{1}{w}_{3}+9{w}_{2}^2)}{3{w}_{2}^2},\\
 & c_{s}^2=\frac{3(2{w}_{1}^2 {w}_2 H-{w}_{2}^2{w}_4+4{w}_1\dot{w_1}{w}_2-2{w}_{1}^2\dot{w_2})}{{w}_1(4{w}_{1}{w}_{3}+9{ w}_{2}^2)},
\end{align}
and
\vspace{-0.5cm}
\begin{align}
\label{w1}
& {w}_1=1-2\delta_D,\\
  \label{W2}
& {w}_2=2H(1-6\delta_D),\\
& {w}_3=-3 H^2(3-\delta_{X}-36\delta_D),\\
&{w}_4=1+2\delta_D.
\end{align}
It is obvious, for defined $\xi$ as a general function of $\phi$, one can easily find the coincidence of the
relations (17)-(22) and their counterparts in \cite{Tsujikawa:2012}. Applying the approximated background
equations in the slow-roll domain (11)-(13), the power spectrum (\ref{Ps}) can be written as
\begin{equation}\label{PsSR}
{\cal P}_{s}\simeq \frac{V^3}{12\pi^2 V_{,\phi}^2}{\cal A}\simeq\frac{V^3}{12\pi^2 V_{,\phi}^2}\Big(1+\theta(\phi)V\Big).
\end{equation}
 The restricted amplitude of the scalar perturbations power spectrum evaluated by the Planck
team from cosmic microwave background (CMB) anisotropy at the CMB pivot scale $(k_{*}=0.05~\rm Mpc^{-1})$ \citep{akrami:2018} is
\begin{equation}\label{psrestriction}
  {\cal P}_{s}(k_{*})\simeq 2.1 \times 10^{-9}.
\end{equation}

Since the scalar spectral index $(n_{s})$ can be calculated from the scaler power spectrum through the definition $n_{s}-1\equiv d\ln{\cal P}_{s}/d\ln k$, therefore one can easily find the following relation between
$n_{s}$ and slow-roll parameters in the NMDC framework \cite{Teimoori:2021}
\begin{align}\label{nsSR}
n_s\simeq  1-\frac{1}{{\cal A}}\left[6\varepsilon_{V}-2\eta_{V}+2\varepsilon_{V}\left(1-\frac{1}{{\cal A}}\right)\right.
 \left.
\times \left(1+\frac{\theta_{,\phi}}{\theta(\phi)}\frac{V(\phi)}{V_{,\phi}}\right)\right],
\end{align}
in which
\begin{equation}\label{eta}
\eta_{V}=\frac{V_{,\phi\phi}}{V}.
\end{equation}
It is evident that for $\theta(\phi)=0$ the coupling term disappears and the relationes of standard slow-roll inflation are recovered.
The slow-roll approximation of  tensor power spectrum at $c_{t}k=aH$ and then the
tensor-to-scalar ratio in NMDC setup have been evaluated in \cite{Tsujikawa:2013} as follows
\vspace{-0.5cm}
\begin{align}\label{PtSR}
&{\cal P}_{t}=\frac{2H^2}{\pi ^{2}},\\
\label{r}
&r\simeq 16 \varepsilon \simeq 16 \frac{\varepsilon_V}{{\cal A}}.
\end{align}
The observational restriction according to the Planck  2018  TT,TE,EE+lowE+lensing+BK14
+BAO data at the 68\%  CL on the scalar spectral index ($n_s$), and the upper bound on the tensor-to-scalar ratio ($r$) at 95\% CL \cite{akrami:2018} are as follows
\begin{align}\label{nsconsraint}
&n_s= 0.9670 \pm 0.0037,\\
\label{rconsraint}
&r<0.065.
\end{align}
\section{Intensification of  Curvature Perturbations  Power Spectrum}\label{sec3}
It is well known that an increase  in the amplitude of curvature perturbations $({\cal R})$ on scales smaller than the CMB scales can lead to  detectable generation of  PBHs and GWs. It is understood that, such a rise in scalar power spectrum can be achieved with the aid of a span of USR inflation. Suitable choices of  parameters and  coupling function between the gravity and  derivatives of  the scalar field in the NMDC setup can  establish such an USR phase due to gravitationally enhanced friction \cite{fu:2019,Teimoori:2021,fu:2020,Dalianis:2020,Heydari:2021}. With this regard,  in this section  we define two-parted form of  coupling function $\theta(\phi)$,  to have a correspondence with the available data on large scales (CMB) and a peak in the curvature perturbations on  smaller  scales, as follows
\begin{equation}\label{t}
\theta(\phi)=\theta_I(\phi)\Big(1+\theta_{II}(\phi)\Big),
\end{equation}
in which
\begin{align}
\label{tI}
&\theta_I(\phi)=\frac{\phi^{\alpha}}{M^{\alpha+2}},\\
\label{tII}
&\theta_{II}(\phi)=\frac{\omega}{\sqrt{\left(\frac{\phi-\phi_c}{\sigma}\right)^2+1}}\,.
\end{align}
The first part of our coupling function (\ref{tI}) is a general form of the ones taken in \cite{Defelice:2011,Granda:2020}. Another part (\ref{tII}) is taken by Fu et all \cite{fu:2019} in order to investigate PBHs and GWs formation in NMDC setup for the potential $V(\phi)\propto\phi^{2/5}$, but we consider a combination of these two functions pursuant to \cite{Dalianis:2020}.  By changing $\alpha$ to $(\alpha-1)$ in (\ref{tI}), regardless of $\alpha$ coefficient, the coupling function of \cite{Dalianis:2020} is retrieved.
Concerning (\ref{tII}), one can easily find that  $\theta_{II}(\phi)$  has a climax at $\phi=\phi_{c}$ of the height and width that are defined by $\omega$ and $\sigma$ respectively. It is obvious that for  the far-off field values of $\phi_{c}$,  the function $\theta_{II}(\phi)$  evanesces. In our model $\theta_{I}(\phi)$ is necessary to make the quantum fluctuations on the CMB scales compatible with the latest data. On the other hand, the quantity $\theta_{I}(\phi)\theta_{II}(\phi)$ depending on the suitable choices for the parameters makes the scalar power spectrum of the curvature perturbations on small scales enlarge to get the enough amplitude to generate the detectable PBHs and GWs. Note that $\alpha$ and $\omega$ are the dimensionless parameters, while $\phi_{c}$, $\sigma$, and $ M$ are parameters with the dimensions of mass.

Note that the quartic potential  is entirely  ruled out by Planck 2018 TT,TE,EE+lowE+len-
sing+BK14+BAO in the standard inflationary groundwork \cite{akrami:2018}. Therefore, this persuades us  to study the quartic potential in the nonminimal derivative coupling framework with the aim of getting compatible observational result with Planck 2018. Hence, we keep on our investigation to see if the quartic inflationary model can derive the observable inflation and generate detectable PBHs and GWs in NMDC framework simultaneously. The quartic potential is given as follows
\begin{equation}\label{v}
V(\phi)=\frac{\lambda}{4}\phi^{4},
\end{equation}
where $\lambda$ is  the dimensionless constant, and   fixed by the scalar  power spectrum restriction (\ref{psrestriction})  in our investigation. From our numerical results listed in Tables \ref{tab1} and \ref{tab2}, one can see for reduced $\lambda$ up to order ${\cal O}(10^{-9})$ in comparison with $\lambda\simeq0.1$ appertain to Higgs field in \cite{Dalianis:2020},  the quartic potential can drive the viable inflation and detectable enhanced PBHs and GWs in the nonminimal derivative coupling groundwork.

Now we want to  estimate the order of needed parameters  to increase the amplitude of the scalar power spectrum (${\cal P}_{s}$) to order ${\cal O}(10^{-2})$,  which is enough value to produce PBHs. Hence, we derive an approximated relation between  ${\cal P}_{s}$ at $\phi=\phi_{c}$ and  ${\cal P}_{s}$ at $\phi=\phi_{*}$,  noting that $\phi_{*}$ is the field value at the time of horizon crossing by $k_{*}$.  So applying Eqs. (\ref{PsSR}) and (\ref{t})-(\ref{v}) we have
 \begin{equation}\label{PsSRHiggs}
   {\cal P}_{s}\simeq \frac{\lambda\ \phi^{6}}{768\ \pi^{2}}\left[1+\frac{1}{4}\ q\ \phi^{4+\alpha} \left(1+\frac{\omega}{\sqrt{1+(\frac{\phi-\phi_{c}}{\sigma})^{2}}}\right)\right],
 \end{equation}
 where
 \begin{equation}\label{q}
   q\equiv\frac{\lambda}{M^{2+\alpha}},
 \end{equation}
 furthermore we suppose that
 \begin{equation}\label{assumptions}
\omega\gg 1, \hspace{1.5cm} \mid \phi_{*}-\phi_{c}\mid\, \gg \sigma \omega ,
\end{equation}
eventually we conclude
\begin{equation}\label{Pspeak}
{\cal P}_{s}\Big|_{\phi= \phi_{c}}\simeq \omega\left(\frac{\phi_{c}}{\phi_{*}}\right)^{10+\alpha}\times{\cal P}_{s}\Big|_{\phi=\phi_{*}}.
\end{equation}
\begin{table}[H]
  \centering
  \caption{The elected parameter assortments  for Cases A,  B,  C,  D, and  F. The value of $\lambda$ is fixed by the power spectrum restriction (\ref{psrestriction}) at horizon crossing $e$-fold number $N_{*}$.}
\begin{tabular}{ccccccc}
  \hline
 $\#$ &\qquad $\omega$\qquad &\qquad$\sigma$\qquad & \qquad$\phi_{c}$\qquad&\qquad $\lambda$\qquad&\qquad $\alpha$\qquad&\qquad $M$\qquad\\[0.5ex] \hline\hline
  Case A& \qquad$5.530\times10^{7}$\qquad &\qquad$2.82\times10^{-11}$\qquad &\qquad$1.3163$ \qquad& \qquad $1.62\times10^{-9}$\qquad& \qquad $16$\qquad& \qquad $0.27$\qquad \\[0.5ex] \hline
  Case B&\qquad $5.594\times10^{7}$\qquad &\qquad$3.79\times10^{-11}$\qquad &\qquad$1.2970$\qquad &\qquad$1.63\times10^{-9}$ \qquad& \qquad $16$\qquad& \qquad $0.27$\qquad  \\ \hline
  Case C&\qquad$6.980\times10^{7}$\qquad &\qquad $6.36\times10^{-11}$\qquad & \qquad$1.2550$\qquad& \qquad$1.89\times10^{-9}$ \qquad& \qquad $16$\qquad& \qquad $0.27$\qquad \\ \hline
  Case D&\qquad$9.762\times10^{7}$\qquad &\qquad $7.30\times10^{-11}$\qquad &\qquad $1.2320$ \qquad&\qquad $1.81\times10^{-9}$\qquad& \qquad $16$\qquad& \qquad $0.27$\qquad  \\ \hline
  Case F&\qquad$6.960\times10^{7}$\qquad &\qquad $4.56\times10^{-11}$\qquad &\qquad $1.1900$ \qquad&\qquad $1.81\times10^{-9}$\qquad& \qquad $24$\qquad& \qquad $0.41$\qquad  \\ \hline
\end{tabular}
 \label{tab1}
\end{table}
\begin{table}[H]
\vspace{-0.6cm}
  \centering
  \caption{The attained outcomes for the Cases of Table \ref{tab1}.  The climax values of the scalar  power spectrum and  fractional abundance of  PBHs  are given by  ${\cal P}_{ s}^\text{peak}$  and   $f_{\text{PBH}}^{\text{peak}}$, respectively.  $M_{\text{PBH}}^{\text{peak}}$ is the PBH mass according to $f_{\text{PBH}}^{\text{peak}}$ . The values of  $n_{s}$ and $r$ are evaluated at horizon crossing CMB $e$-fold number $(N_{*})$.}
\begin{tabular}{ccccccc}
  \hline
   $\#$ & \qquad $n_{s}$\qquad &\qquad $r$\qquad &\qquad$ {\cal P}_{s}^{\text{peak}}$\qquad &\qquad$k_{\text{peak}}/\text{Mpc}^{-1}$\qquad& \qquad$f_{\text{PBH}}^{\text{peak}}$\qquad& \qquad$M_{\text{PBH}}^{\text{peak}}/M_{\odot}\qquad$\\ \hline\hline
  Case A &\qquad0.9730\qquad  &\qquad0.0436\qquad& \qquad0.050\qquad & \qquad$2.89\times10^{6}$ \qquad&\qquad0.0010\qquad &\qquad$28.23$\qquad \\ \hline
 Case B &\qquad 0.9668\qquad  & \qquad0.0440\qquad &\qquad0.042 \qquad&\qquad  $6.91\times10^{8}$ \qquad&\qquad0.0434\qquad &\qquad $4.97\times10^{-6}$ \qquad\\ \hline
 Case C &\qquad0.9633\qquad  & \qquad0.0468\qquad & \qquad0.034\qquad & \qquad$2.86\times10^{12}$\qquad &\qquad 0.9831\qquad &\qquad$2.88\times10^{-13}$ \qquad\\ \hline
 Case D &\qquad 0.9615\qquad  &\qquad0.0489\qquad &\qquad0.030\qquad & \qquad$3.94\times10^{13}$\qquad
 &\qquad 0.9911\qquad&\qquad$1.52\times10^{-15}$\qquad \\ \hline
 Case F &\qquad 0.9636\qquad  &\qquad 0.0356\qquad &\qquad0.035\qquad & \qquad$1.03\times10^{12}$ \qquad& \qquad 1 \qquad&\qquad$2.21\times10^{-12}$\qquad \\ \hline
\end{tabular}
\label{tab2}
\end{table}
\noindent
Clearly, regarding (\ref{psrestriction}) and (\ref{Pspeak}), the climax of the power spectrum at $\phi=\phi_{c}$ will be of order ${\cal O}(10^{-2})$ provided by $\omega\sim{\cal O}(10^{7})$.
We know that in the vicinity of peak position the term of $\theta_{I}(\phi)\theta_{II}(\phi)$ is dominated in (\ref{t}), ergo  the power spectrum (\ref{PsSR}) around the peak position abbreviates as the following form
\begin{align}\label{Psaround peak}
{\cal P}_{s}\big|_{\phi\sim \phi_{c}}\simeq \frac{\lambda\ \phi^{6}}{768\ \pi^{2}}\left[1+\frac{\ q\ \phi^{4+\alpha}}{4} \left( \frac{\omega}{\sqrt{1+(\frac{\phi-\phi_{c}}{\sigma})^{2}}}\right) \right]
\simeq {\cal B}\phi^{10+\alpha}\left( \frac{\omega}{\sqrt{1+(\frac{\phi-\phi_{c}}{\sigma})^{2}}}\right),
\end{align}
where
\begin{align}\label{B}\nonumber
 {\cal B}\equiv\frac{\lambda\  q}{3072\ \pi^{2}}.
\end{align}
It is obvious from (\ref{Psaround peak}) that, the width of ${\cal P}_{s}$ around the peak position is determined by parameter $\sigma$.
In view of these estimations and expressed conditions in (\ref{assumptions}),  we set  $q=20$  and  contemplate five different Cases of parameters  which are represented in Table \ref{tab1}. It is worth noting that an assortment consists of seven free parameters  $\{\alpha, M,   \lambda,  q, \omega, \phi_c, \sigma\}$ identifies our model, in which the parameters $\alpha$, $M$, $q$, and $\lambda$ are correlated together via Eq. (\ref{q}).  The  calculated  values of inflationary observables  $n_{s}$, $r$, and  quantities related to generated PBHs are represented in Table \ref{tab2}.

\begin{figure*}
\begin{minipage}[b]{1\textwidth}
\vspace{-1cm}
\subfigure{\includegraphics[width=.45\textwidth]%
{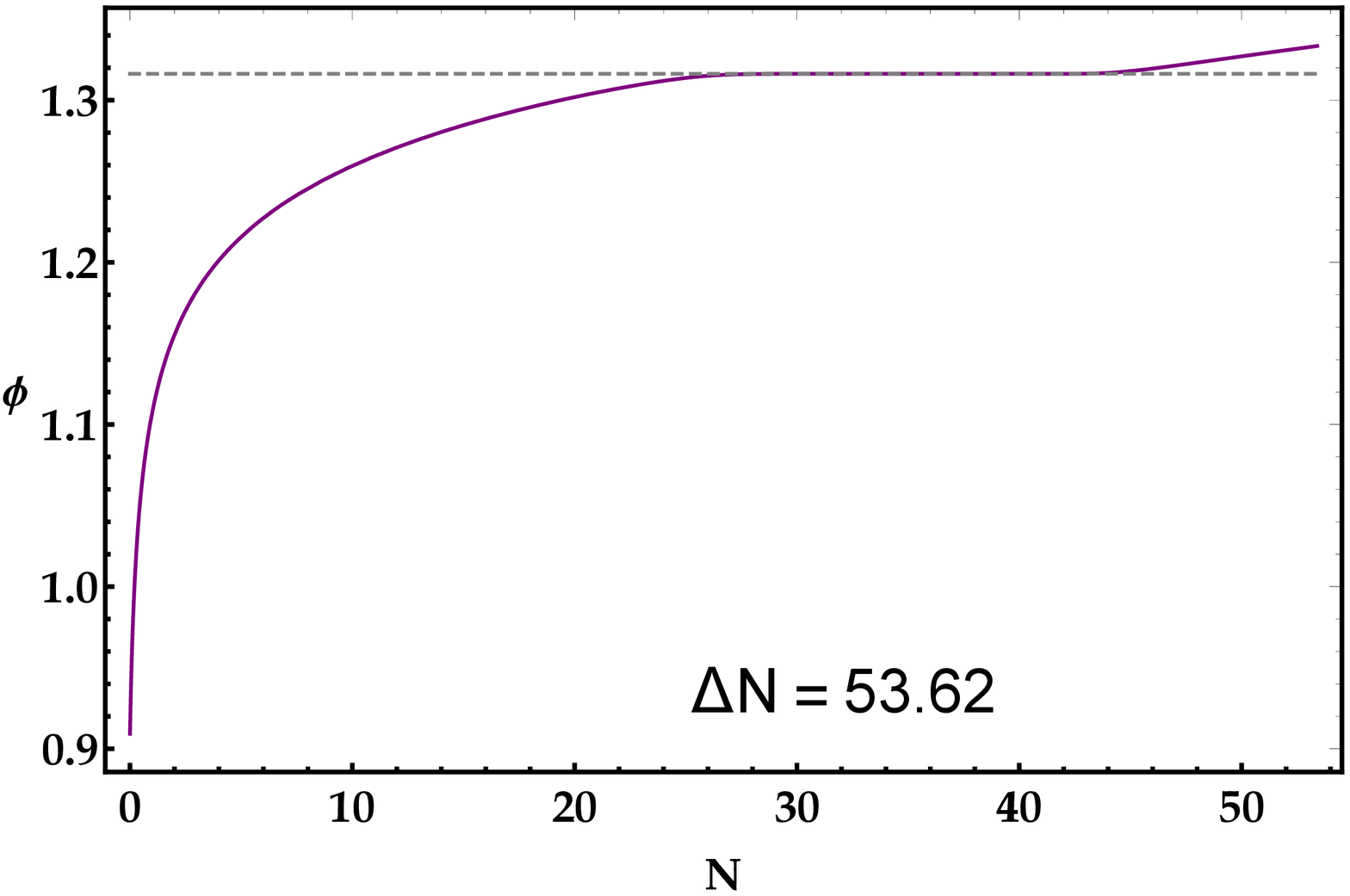}} \hspace{.1cm}
\subfigure{ \includegraphics[width=.45\textwidth]%
{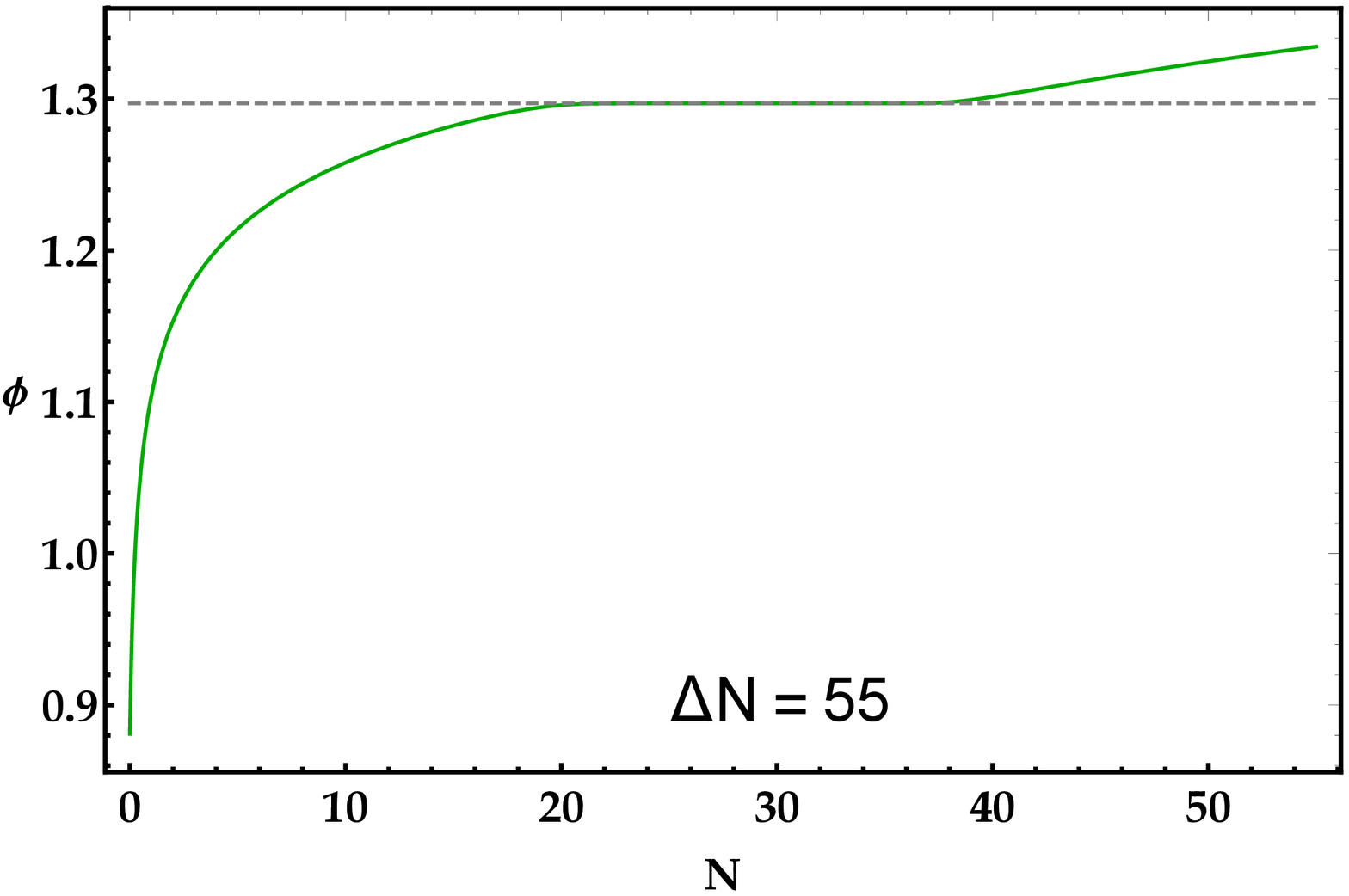}}\hspace{.1cm}
\subfigure{ \includegraphics[width=.45\textwidth]%
{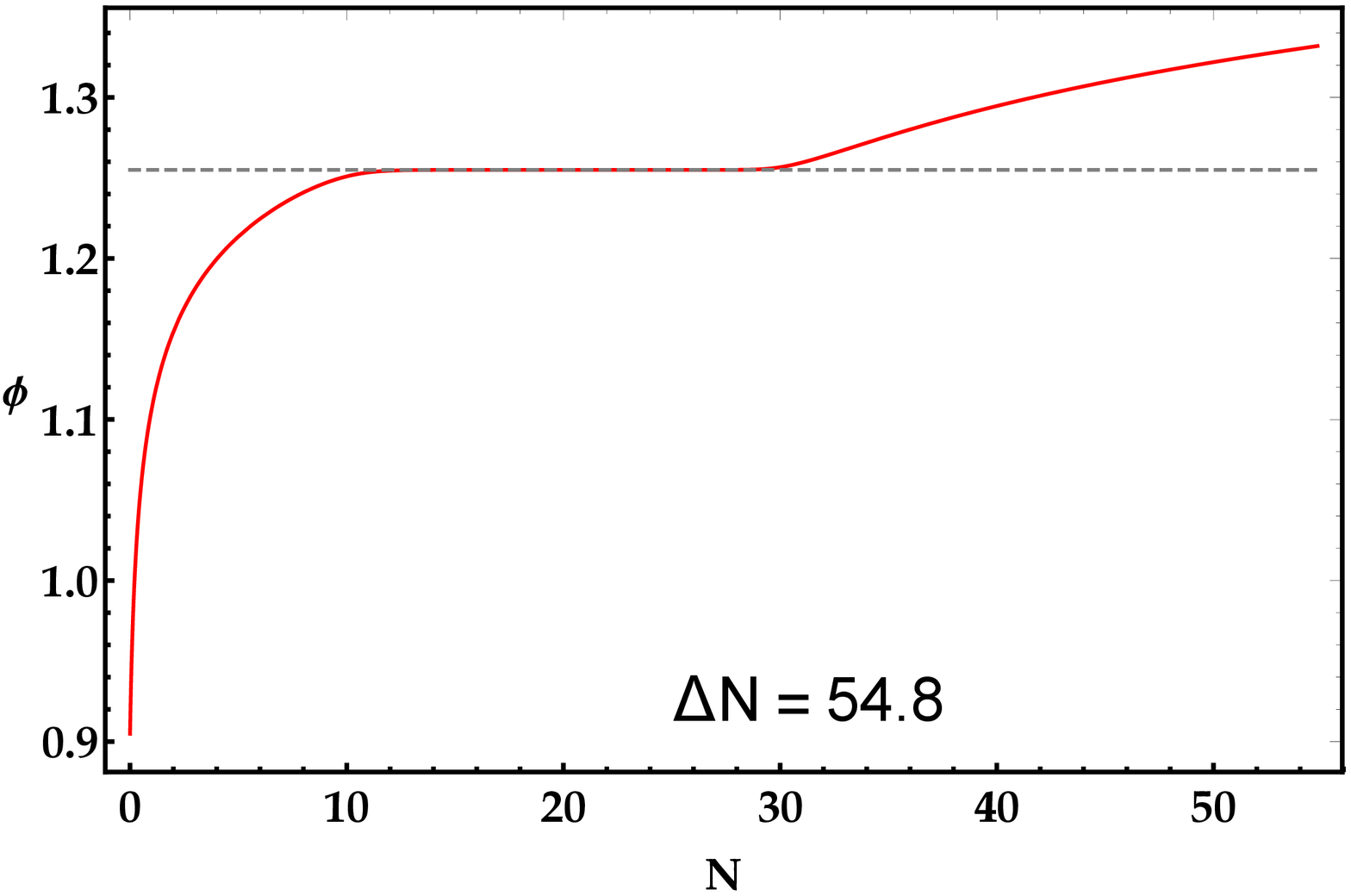}}\hspace{.1cm}
\subfigure{ \includegraphics[width=.45\textwidth]%
{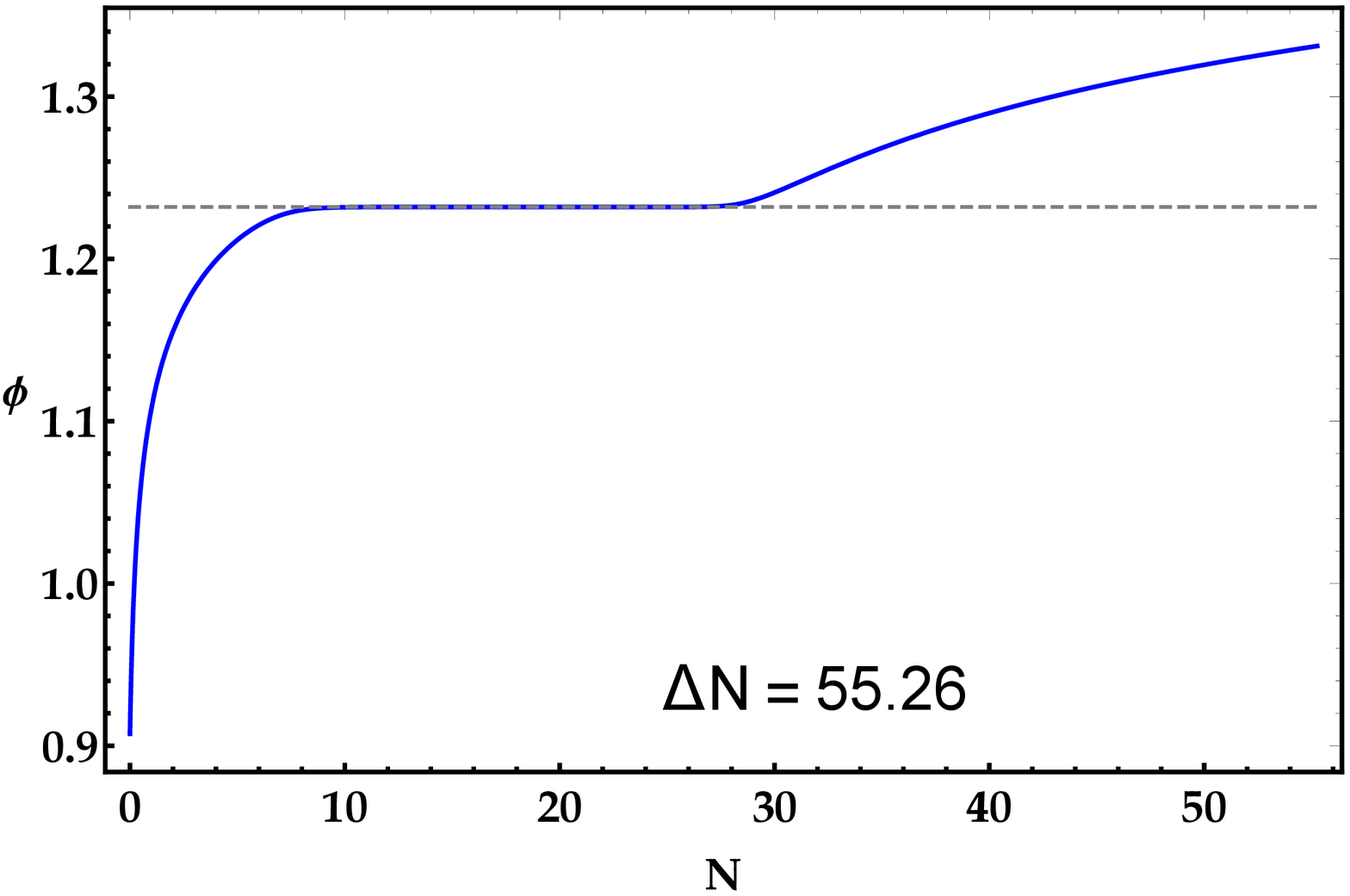}}
\subfigure{ \includegraphics[width=.45\textwidth]%
{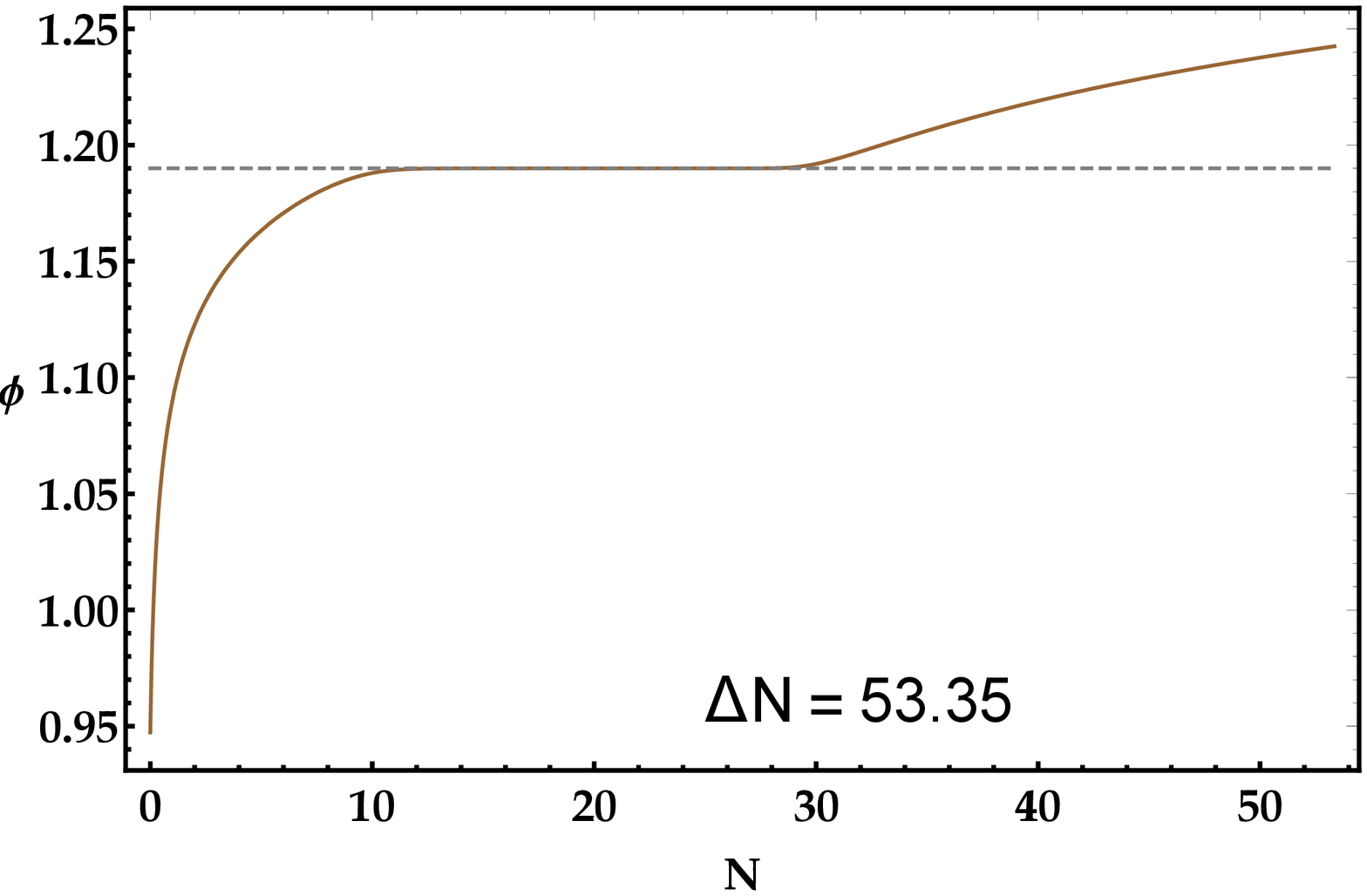}}
\end{minipage}
\vspace{-1cm}
\caption{Evolution of the scalar field $\phi$ as a function of  $e$-fold number $N$ for Case A (purple line), Case B (green line),  Case C (red line), Case D (blue line), and Case F (brown line). The dashed lines show $\phi=\phi_{c}$ for each Case and  the initial conditions are found by using Eqs. (\ref{FR1:SRC}) and (\ref{Field:SRC}).}
\label{fig:phi}
\end{figure*}

\begin{figure*}
\begin{minipage}[b]{1\textwidth}
\vspace{-2.cm}
\subfigure{\includegraphics[width=.40\textwidth]%
{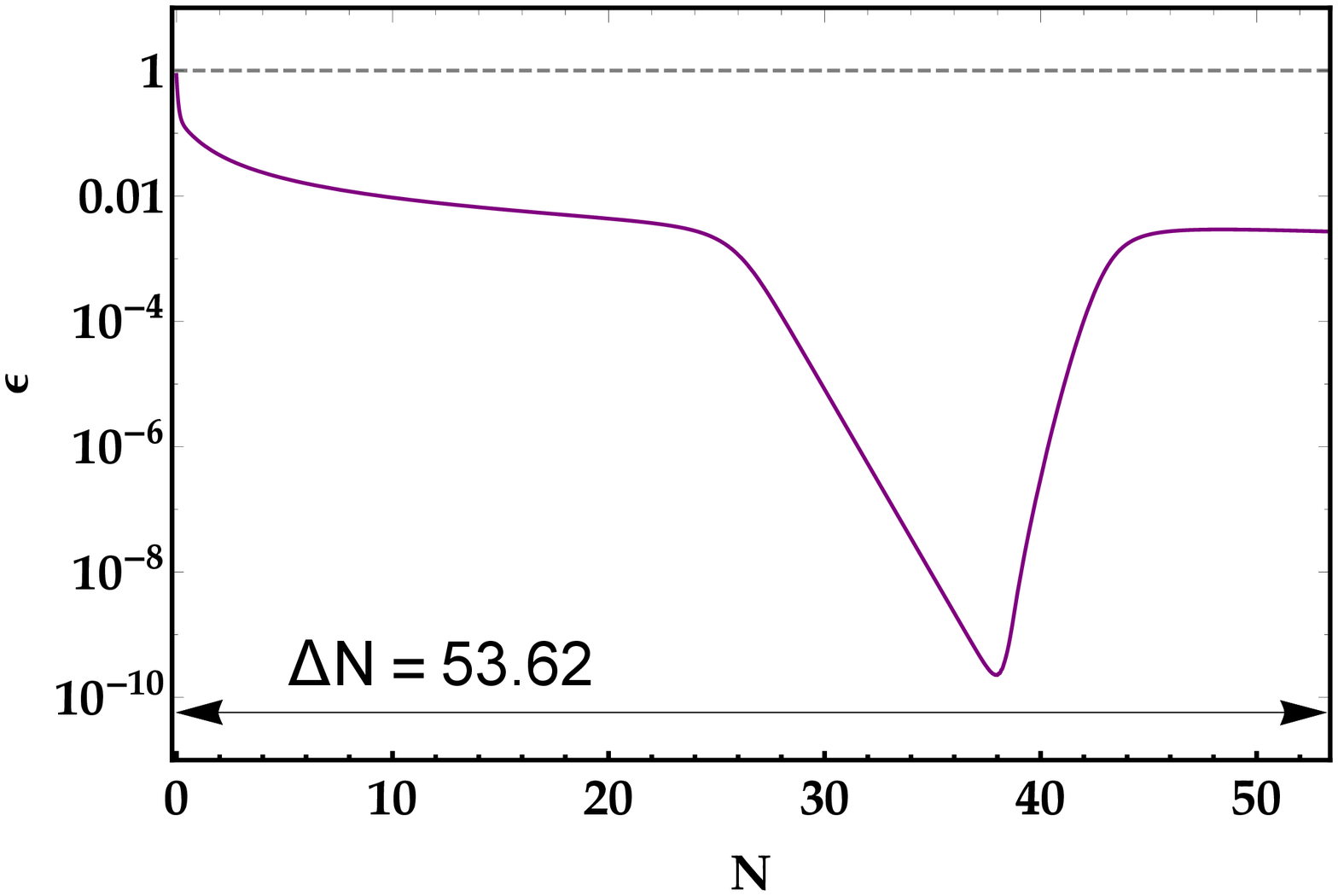}} \hspace{.09cm}
\subfigure{ \includegraphics[width=.40\textwidth]%
{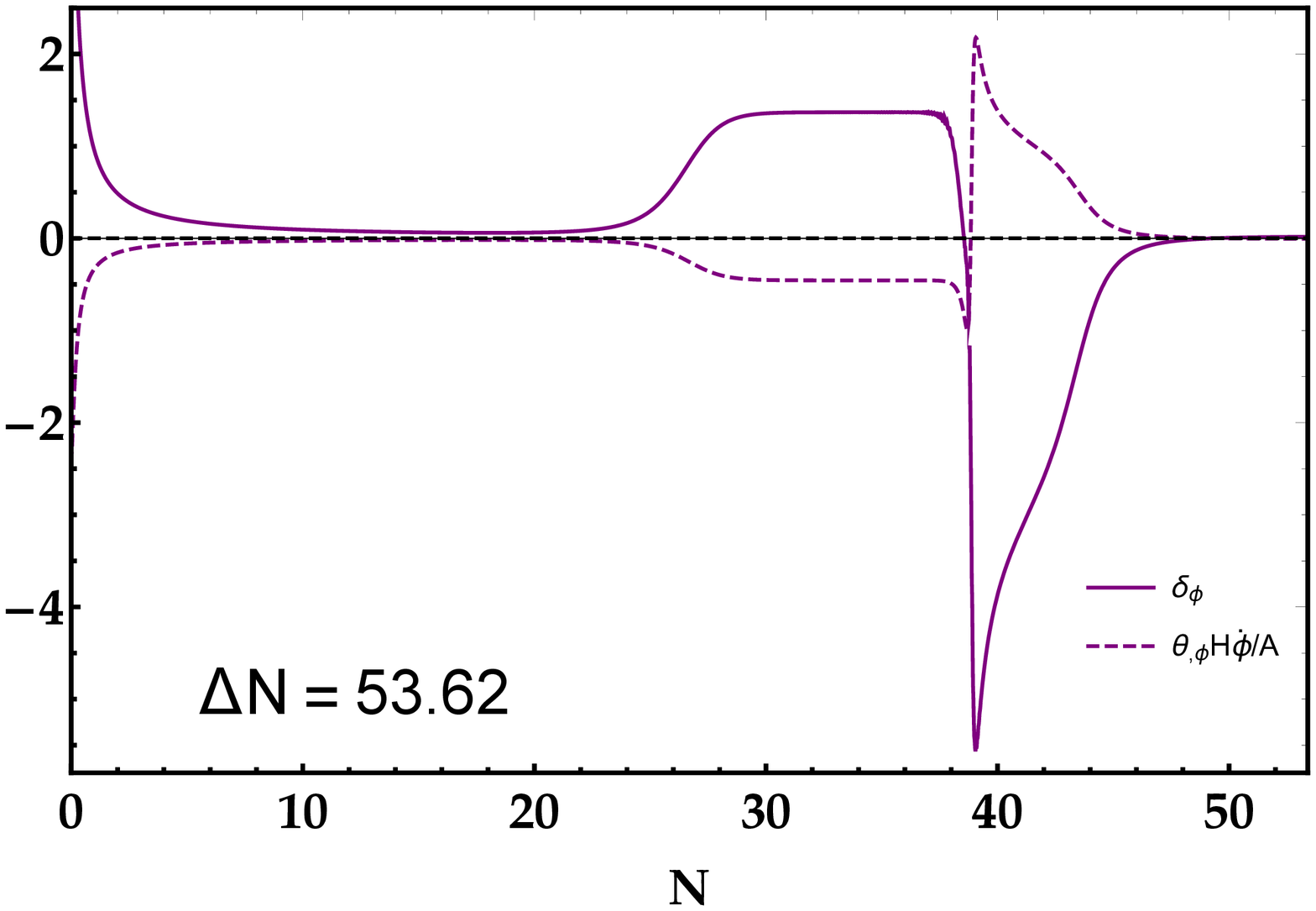}}\hspace{.09cm}
\subfigure{ \includegraphics[width=.40\textwidth]%
{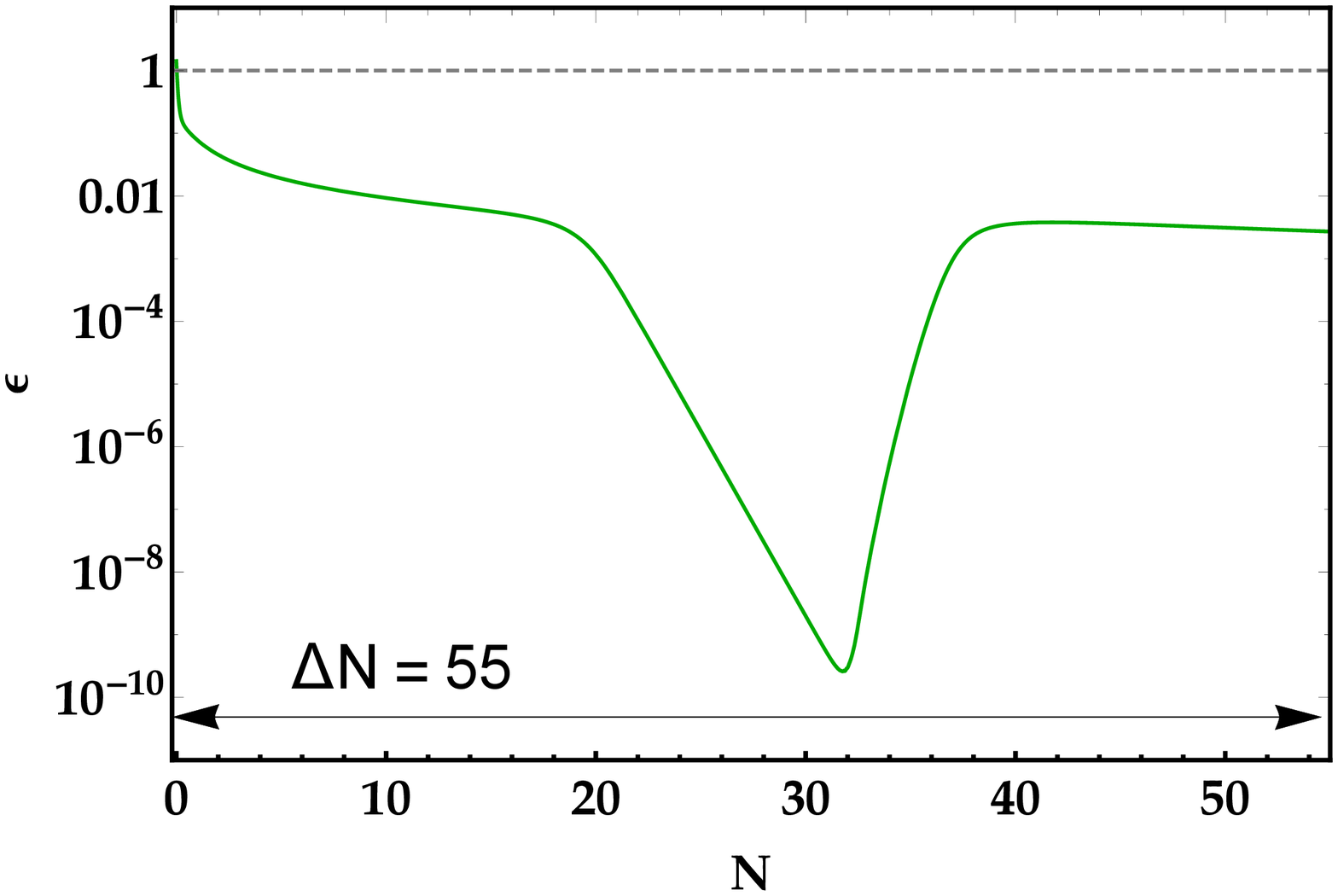}}\hspace{.09cm}
\subfigure{ \includegraphics[width=.40\textwidth]%
{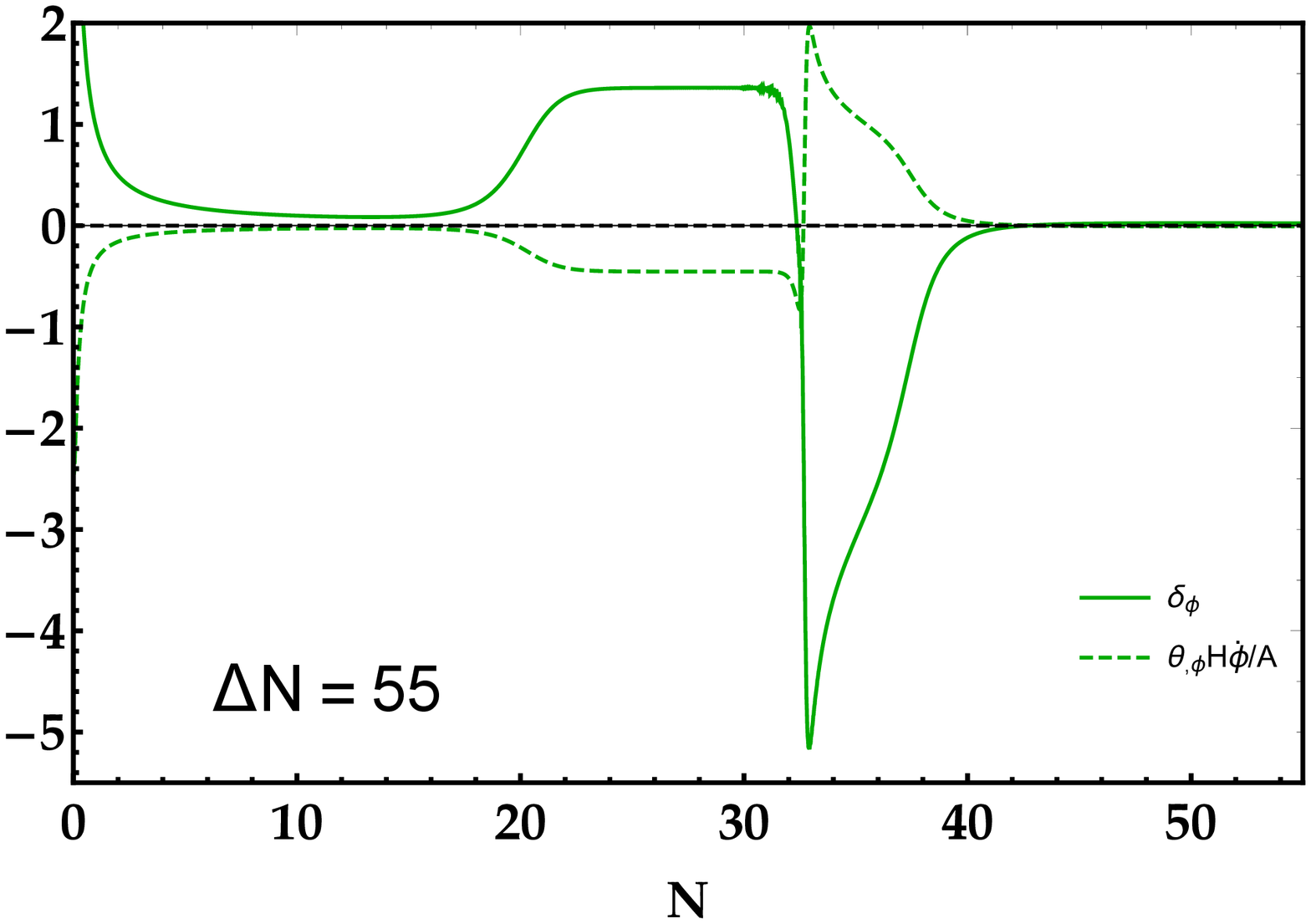}} \hspace{.09cm}
\subfigure{\includegraphics[width=.40\textwidth]%
{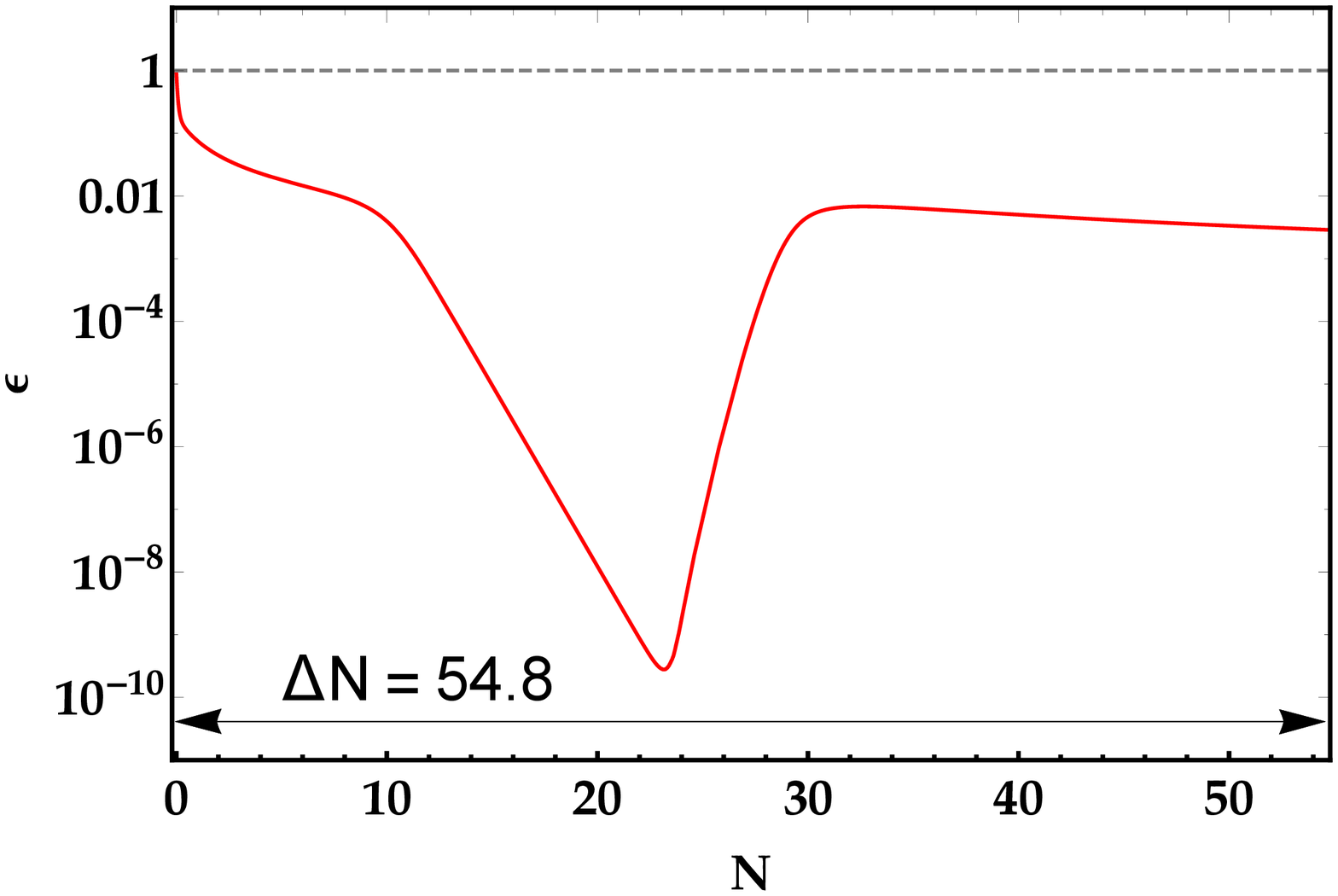}} \hspace{.09cm}
\subfigure{ \includegraphics[width=.40\textwidth]%
{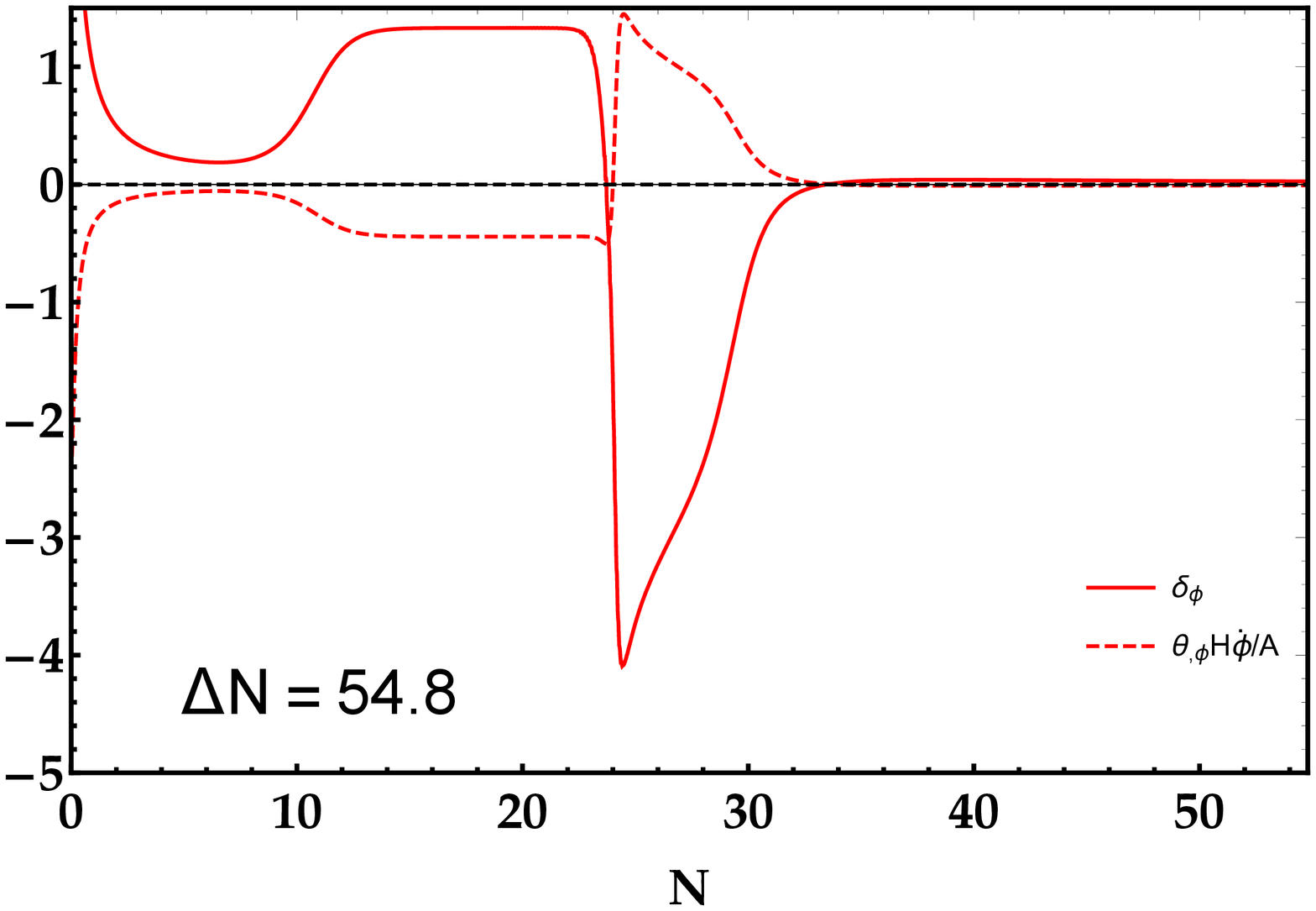}}\hspace{.09cm}
\subfigure{ \includegraphics[width=.40\textwidth]%
{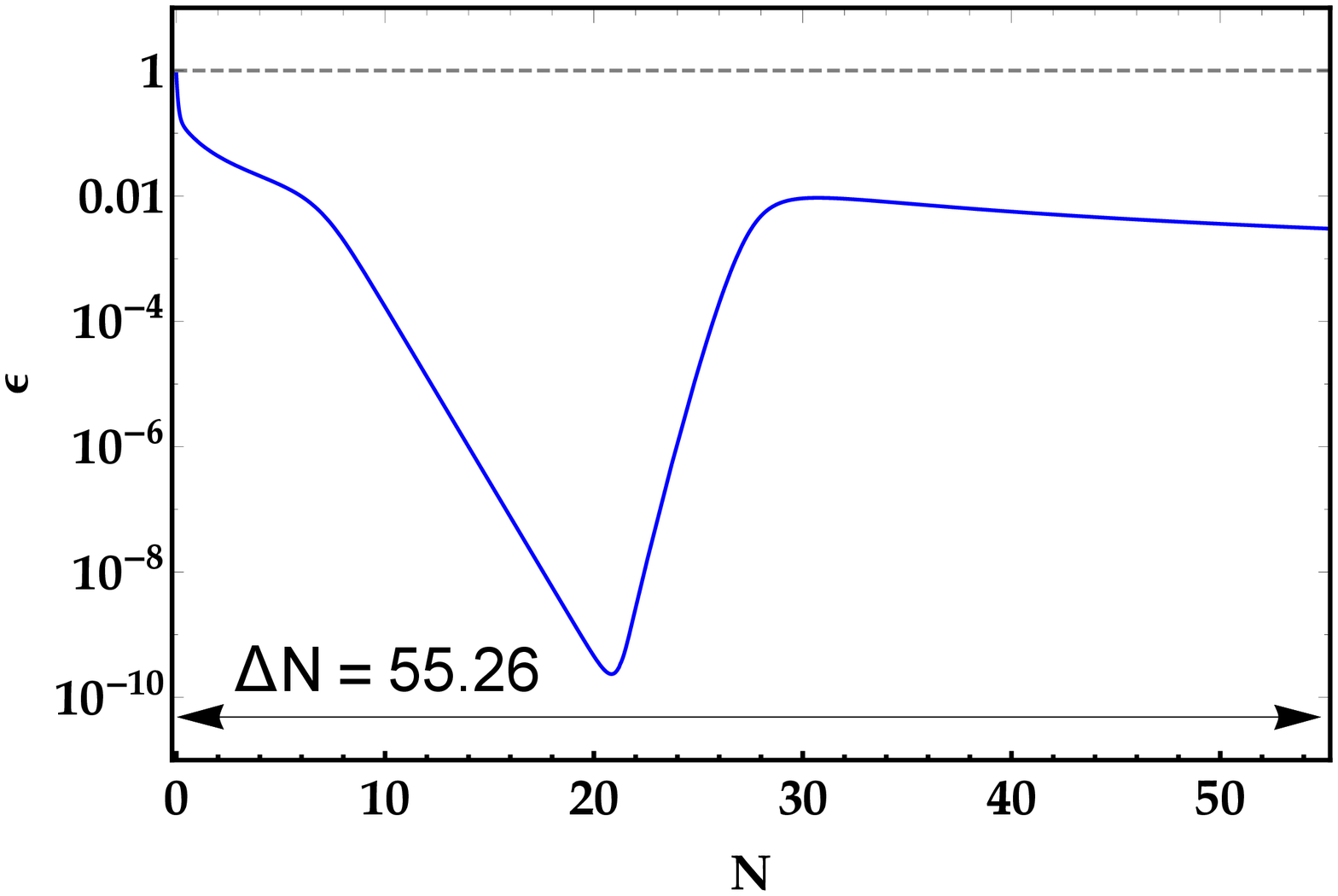}}\hspace{.09cm}
\subfigure{ \includegraphics[width=.40\textwidth]%
{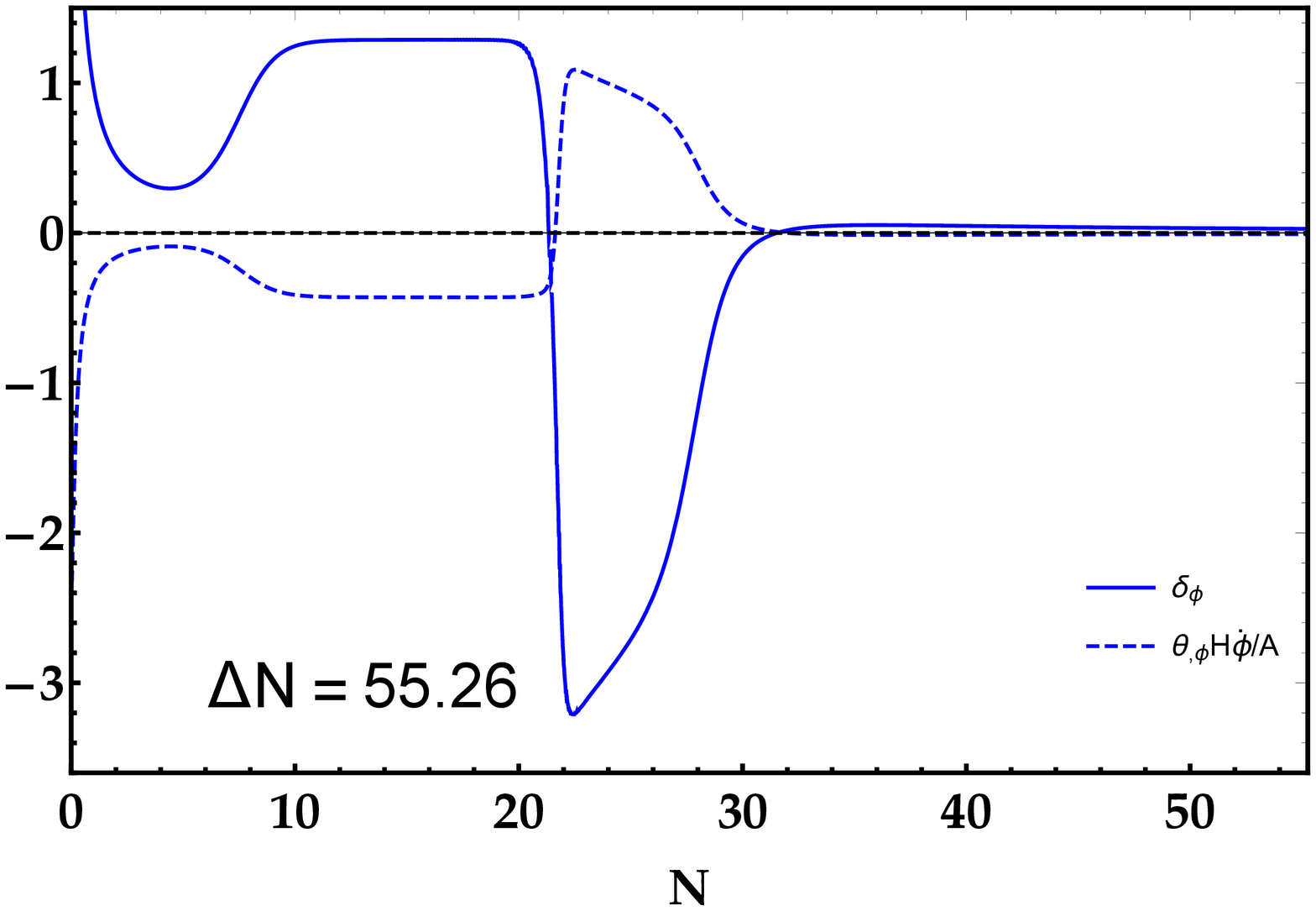}}\hspace{.09cm}
\subfigure{ \includegraphics[width=.40\textwidth]%
{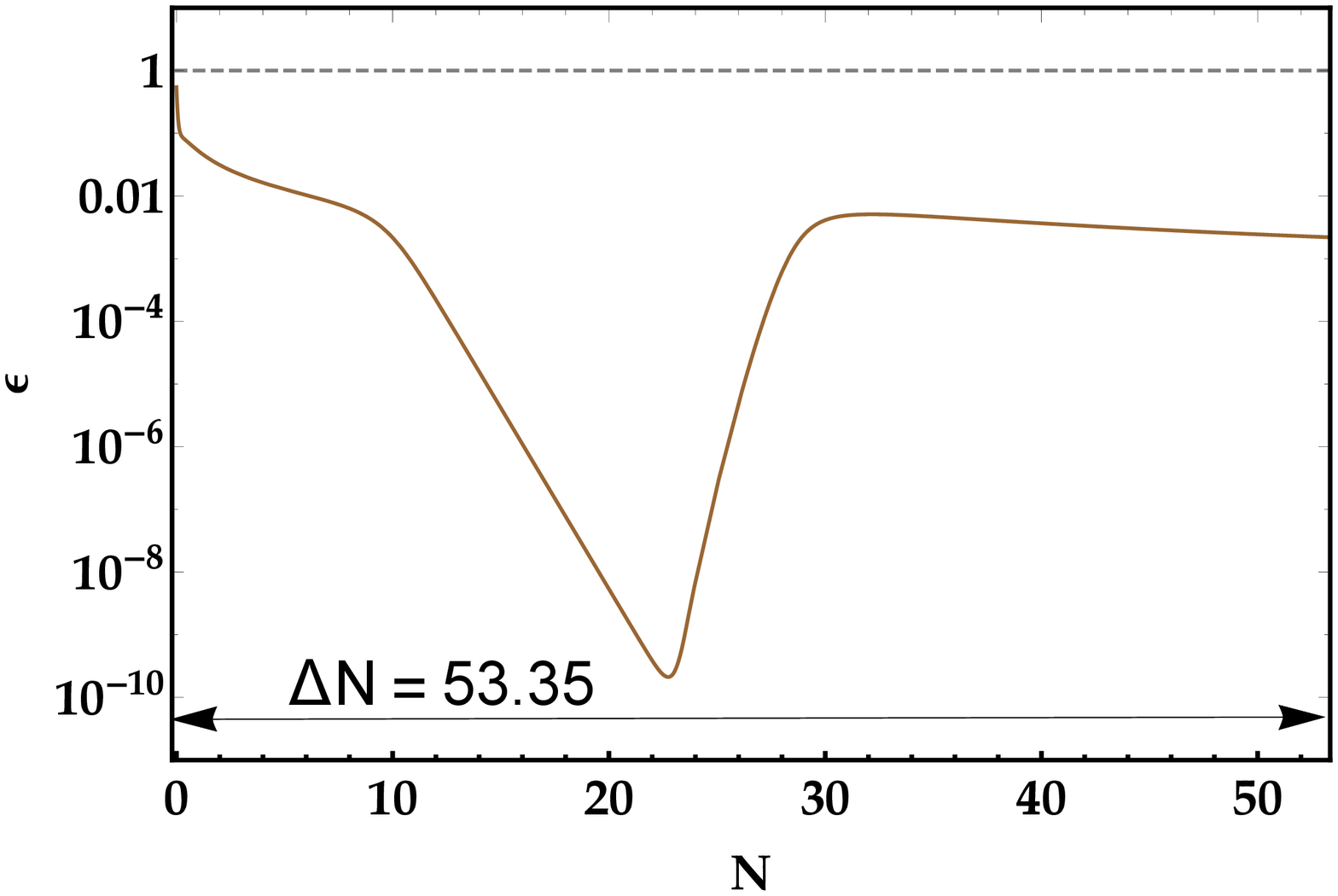}}\hspace{.09cm}
\subfigure{ \includegraphics[width=.40\textwidth]%
{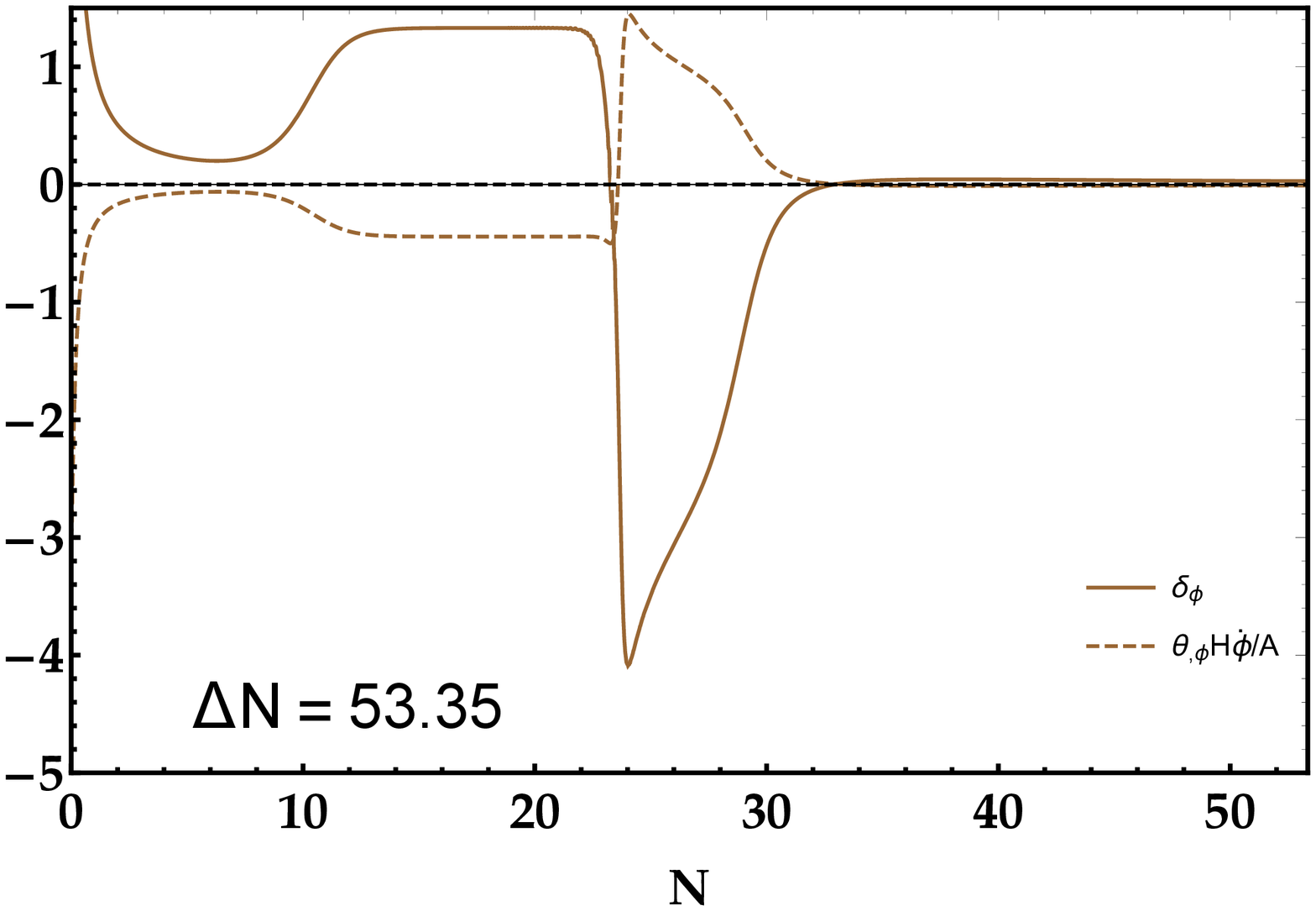}}
\end{minipage}
\vspace{-1.3cm}
\caption{(left) Evolution of the first slow-roll parameter $\varepsilon$, (right) the second slow-roll parameter $\delta_{\phi}$, and  ${\theta_{,\phi}H\dot{\phi}}/{\cal A}$ with respect to the $e$-fold number $N$ for Case A (purple line), Case B (green line),  Case C (red line), Case D (blue line), and Case F (brown line).}
\label{fig:SRp}
\end{figure*}

\begin{figure}[H]
\centering
\vspace{-0.3cm}
\scalebox{0.6}[0.6]{\includegraphics{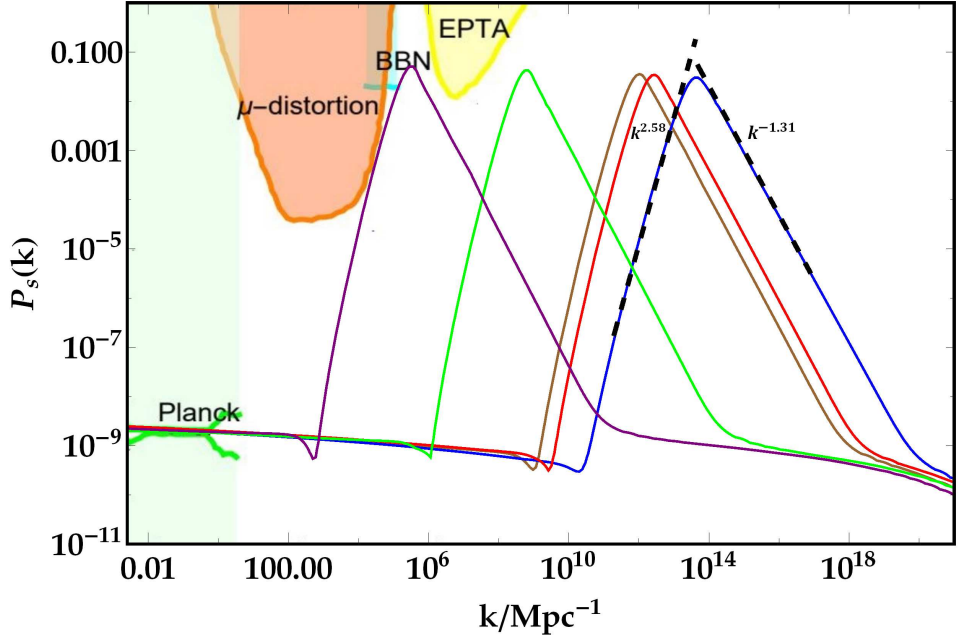}}
\vspace{-0.6cm}
\caption{The  obtained exact scalar power spectrum via resolving the Mukhanov-Sasaki equation  in terms of comowing  wavenumber $k$. The purple, green, red, blue, and brown lines are according to the Cases A, B, C, D, and F respectively.  The light-green shadowy zone represents the CMB observations \cite{akrami:2018}. The yellow, cyan, and orange shadowy zones  represent the restrictions from the PTA observations \cite{Inomata:2019-a},   the effect on the ratio between neutron and proton during the big bang nucleosynthesis (BBN) \cite{Inomata:2016}, and the $\mu$-distortion of CMB \cite{Fixsen:1996} respectively. The  power-law form of ${\cal P}_{s}$ with respect to $k$ is illustrated  by black dashed line for Case D.}
\label{fig-ps}
\end{figure}

It is known that  the observable inflation takes place in about 50-60 $e$-folds number from the horizon crossing moment by $k_{*}$ to the end of inflation. So regarding the reheating process duration, we set the horizon crossing $e$-fold number ($N_{*}$) as $53.62$ for Case A, $55$ for Case B, $54.8$  for Case C,  $55.26$ for Case D, and $53.35$ for Case F. From Fig. \ref{fig:SRp}, one can see that the end of inflation epoch is specified by resolving $\varepsilon =1$ at the $e$-fold number $N_{\text{end}}=0$ for each Case.
We solve the background equations (\ref{FR1:eq})-(\ref{Field:eq}), using the quartic potential (\ref{v}) and the NMDC coupling function (\ref{t})-(\ref{tII}), numerically. Then we represent the exact behavior of the scalar field  $\phi$ with respect to  the  $e$-fold number $N$, when $dN=-Hdt$,  from the horizon crossing ($N_{*}$) to the end of inflation ($N_{\text{end}}$) in Fig. \ref{fig:phi}, for Case A (purple line), Case B (green line), Case C (red line),  Case D (blue line), and Case F (brown line). In this figure, one can see a plane zone lasting about 20 $e$-folds  around $\phi=\phi_{c}$ in all field excursion. In this zone  the scalar field revolves very slowly owing to the high friction, and the slow-roll condition is broken, thus one has an Ultra Slow-Roll (USR) stage. Figure \ref{fig-ps} shows that, the curvature perturbations power spectrum will be intensified  during this USR stage.
Figure \ref{fig:SRp} shows the evolution of the  first slow-roll parameter $(\varepsilon)$ (left plot),  the second slow-roll parameter $(\delta_{\phi})$, and  ${\theta_{,\phi}H\dot{\phi}}/{\cal A}$ (right plot) with respect to $e$-fold number ($N$) during the observable inflation ($\Delta N=N_{*}-N_{\text{end}}$) for Cases listed in Table \ref{tab1}. In this figure a severe reduction in the value of  $\varepsilon$ in the USR stage for each Case  up to the order ${\cal O}(10^{-10})$ can be seen, which gives rise to a large amplification in the amplitude of the  curvature power spectrum.   Also  one can see $\delta_{\phi}$ violates the slow-roll condition due to exceed unity in this stage temporarily. As regards the evolution of  the slow-roll parameters at horizon crossing $e$-fold number ($N_{*}$), one can see that the  previously mentioned slow-roll conditions
are respected for all Cases,  therefore we can use Eqs. (\ref{nsSR}) and (\ref{r}) to evaluate the scalar spectral index ($n_{s}$) and tensor-to-scalar ratio ($r$) in our model.
Regarding the numerical outcomes  listed in  Table \ref{tab2}, we can realize that, the values of $r$ for all Cases and $n_{s}$ for Cases B and F are consonant with
the $68\%$ CL of the Planck  2018  TT,TE,EE+lowE+lensing+BK14+BAO data, and the values of $n_{s }$ for Cases A, C, and  D are  consistent  with $95\%$  CL \cite{akrami:2018}.
Consequently, we have been able to revive the ruled out quartic potential in the standard inflationary scenario by means of selecting the nonminimal derivative coupling framework.
It is worth mentioning  that, the recent BICEP/Keck collaboration's constraints suggest that the tensor-scalar ratio  $r<0.036$ \cite{BICEP:2021}, which thereunder as we can see for Case F of Tables \ref{tab1} and \ref{tab2} the parameter $\alpha\geq24$ should have been chosen  for our model.

Whereas Eq. (\ref{PsSRHiggs}) is acquired under the slow-roll conditions with the proviso (\ref{condition}), applying that to calculate the scalar power spectrum in the USR stage is impermissible owing to infraction of the slow-roll limitation in this stage. Inevitably, to calculate the exact scalar power spectrum,  numerical analysis of the following Mukhanov-Sasaki (MS) equation is
needed, for all Fourier modes
\begin{equation}\label{MS}
 \upsilon^{\prime\prime}_{k}+\left(c_{s}^2 k^2-\frac{z^{\prime\prime}}{z}\right)\upsilon_k=0.
\end{equation}
The MS equation (\ref{MS}) represents the  evolution of curvature perturbations (${\cal R}$) in the Fourier space ($\upsilon_k$) during inflation epoch, from sub-horizon scale ($c_{s}k\gg aH$) until  exiting  the
horizon and getting the constant value at super-horizon scales  ($c_{s} k\ll aH$).  In (\ref{MS}) the prime indicates a derivative with respect to the conformal time $\eta=\int {a^{-1}dt}$, and
\begin{equation}\label{z}
 \upsilon\equiv z {\cal R}, \hspace{1cm} z=a\sqrt{2Q_s}.
\end{equation}
We consider the following  Fourier transform of  the  Bunch-Davies vacuum state as the initial condition for the sub-horizon scale ($c_{s}k\gg aH$) \cite{Defelice:2013}
\begin{equation}\label{Bunch}
\upsilon_k\rightarrow\frac{e^{-i c_{s}k\eta}}{\sqrt{2c_s k}}.
\end{equation}
After finding the numerical solutions of  the  Mukhanov-Sasaki equation (\ref{MS}),  the exact scalar power-spectrum for  each mode $\upsilon_k$ is obtained as
\begin{equation}\label{PsBunch}
{\cal P}_{s}=\frac{k^3}{2\pi^2}\Big|{\frac{{\upsilon_k}^2}{z^2}}\Big|_{c_{s}k\ll aH}.
\end{equation}
\noindent
The numerical results for the exact value of the climax of the scalar power spectrum (${\cal P}_{s}^{\rm peak}$), and the related comoving wavenumber $(k_{\rm peak})$ for each Case of Table \ref{tab1} are represented in Table \ref{tab2}. Also, in Fig. \ref{fig-ps} the exact power spectra for all Cases of  Table \ref{tab1} as a function of comoving wavenumber ($k$), and the current observational restrictions are illustrated. The purple, green, red,  blue, and brown  lines are related to the actual  ${\cal P}_{s}$ for Cases A, B, C,  D, and F, respectively. This figure exhibits that, during the slow-roll inflationary domain on large scales around the CMB scale ($k\sim0.05~ \rm Mpc^{-1}$), the  power spectra for all Cases are of order  ${\cal O}(10^{-9})$ which are consonant with the observational data (\ref{psrestriction}). Furthermore,  in the USR phase on smaller scales, one can see the multiplication of the power spectra to  order  ${\cal O}(10^{-2})$, which is sufficient amount to generate PBHs.
Into the bargain, we have inferred that the scaling behaviour of ${\cal P}_{s}$  in the vicinity of  peak position can be
\noindent
 appraised as power-law function  versus  wavenumber  ${\cal P}_{s}\sim k^{n}$. Thereby for Case D, ${\cal P}_{s}\sim k^{2.58}$ for $k<k_{\rm peak}=3.94\times 10^{13}\ {\rm Mpc^{-1}}$, and ${\cal P}_{s}\sim k^{-1.31}$ for $k>k_{\rm peak}=3.94\times 10^{13}\ {\rm Mpc^{-1}}$ have been delineated in Fig. \ref{fig-ps}.
\section{The reheating stage}\label{sec4}
To keep on our study on the produced  PBHs and GWs from the intensified scalar power spectrum, we should know when the super-horizon perturbation modes come back to the horizon after the inflation epoch to collapse. It is known that to shift from a supercooled universe after the inflation epoch to the hot RD epoch a thermalizing phase, so-called reheating stage, is required (see \cite{Allahverdi:2010} for a review). In spit of  uncertainty of the physics of reheating, one can use the simple canonical outline \cite{Abbot:2010,Dolgov:2010,Albrecht:2010}, whereby generated cold gases from the oscillations of the inflaton filed around the minimum of the potential transform to a plasma of  relativistic particles, after that the standard RD area of the Hot Big Bang theory takes over. The effective equation-of-state parameter of the canonical reheating stage is $\omega_{\rm reh}=0$, and  duration of this epoch is proportional to the inflaton decay-rate and reheating temperature ($T_{\rm reh}$). One should attach the importance of the reheating consideration insomuch that prolonged reheating stage with the low temperature can result in  return   the climax scales of the curvature power spectrum to the horizon in the reheating stage and  form the PBHs in this stage instead of RD epoch. Accordingly, in the following we try to find a  criterion to show that in our investigation the mentioned scales  come back to the horizon in RD epoch. In this regard, we pursue the used technique in \cite{Dalianis:2019,mahbub:2020}.

Suppose that a scale like $k^{-1}$ exits the horizon $\Delta N_{k}$ $e$-folds before the end of inflation, right at the moment  it  returns to the horizon we have
\begin{equation}\label{deltank}
\left( \frac{a_{k,\text{re}}}{a_{\text{end}}} \right)^{\frac{1}{2}(1+3w)}=e^{\Delta N_{k}},
\end{equation}
wherein  $a_{k,\text{re}}$, $a_{\text {end}}$, and $w$  denote the scale factor at the horizon re-entry,  the scale factor at the end of inflation,  and the equation-of-state parameter, respectively. Then we define the post-inflationary $e$-folds number related to the scale $k^{-1}$ from   the time of re-entry to the end of inflation as $\tilde{N}_{k}$,
\begin{equation}\label{nk}
\tilde{N}_{k}\equiv \ln\left( \frac{a_{k,\text{re}}}{a_{\text{end}}} \right).
\end{equation}
Thus,  from Eqs. (\ref{deltank}) and (\ref{nk}) we drive the following relation
\begin{equation}\label{deltaNtoN}
\tilde{N}_{k}=\frac{2}{1+3w}\Delta N_{k}.
\end{equation}
Duration of the  reheating stage is defined by the $e$-folds number from the end of reheating to the end of inflation as $\tilde{N}_{\text{reh}}\equiv \ln\left( a_{\text{reh}}/a_{\text{end}} \right)$,
wherein  $a_{\rm reh}$ denotes the scale factor at the end of reheating. Applying the continuity equation, we can specify the $e$-folds number $\tilde{N}_{\text{reh}}$ in terms of the energy density at the end of reheating stage $(\rho_{\text{reh}})$ as follows
\begin{equation}\label{nktilde}
\tilde{N}_{\rm reh}=\frac{1}{3(1+w_{\text{reh}})}\ln\left( \frac{\rho_{\text{end}}}{\rho_{\text{reh}}} \right),
\end{equation}
 in which the energy density at the end of inflation is evaluated via $\rho_{\text{end}}=3H_{\text{end}}^{2}$.
By making a comparison between  $\tilde{N}_{k}$ and $\tilde{N}_{\text{reh}}$ one can specify when the  horizon re-entry takes place for the scale  $k^{-1}$ associated with the PBH formation. In the way that,
For $\tilde{N}_{k}>\tilde{N}_{\text{reh}}$, the horizon re-entry takes place during the RD era, and for
$\tilde{N}_{k}<\tilde{N}_{\text{reh}}$ the mentioned scale comes back to the horizon during the reheating stage.
\begin{table}[H]
\centering
\caption{The values of $N_{ k}^{\text{peak}}$, $\Delta N_{ k}^{\text{peak}}$, $\tilde{N}_{\text{reh}}$,  $k_{\text{reh}}$, $T_{\text{reh}}$, and  $\Delta N_{\text{peak}}^{(\text{cr})}$   for all Cases of Table \ref{tab1}.}
\begin{tabular}{cccccccc}
  \hline
   $\#$ &\qquad$\Delta N$\qquad & \qquad$N_{k}^{\text{peak}}$\qquad&\qquad$\Delta N_{k}^{\text{peak}}$\qquad &\qquad$\tilde{N}_{\text{reh}}$\qquad&\qquad$k_{\rm reh}/\rm Mpc^{-1}$\qquad & \qquad$T_{\text{reh}}/\rm GeV$\qquad&\qquad$\Delta N_{\text{peak}}^{(\text{cr})}$\qquad \\ \hline\hline
  Case A &\qquad53.62\qquad &\qquad38\qquad &\qquad15.62\qquad& \qquad10.9\qquad&\qquad$2.20\times10^{19}$\qquad &\qquad$1\times10^{12}$\qquad &\qquad12.2\qquad \\ \hline
  Case B & \qquad55\qquad &\qquad 31.6\qquad &\qquad 23.4\qquad &\qquad5.1\qquad&\qquad$1.82\times10^{21}$\qquad &\qquad$1\times10^{14}$ \qquad & \qquad16.4\qquad \\ \hline
  Case C &\qquad 54.8 \qquad&\qquad23\qquad &\qquad 31.8\qquad &\qquad5.9\qquad&\qquad$1.04\times10^{21}$\qquad &\qquad$6\times10^{13}$\qquad  &\qquad22.1\qquad \\ \hline
  Case D &\qquad 55.26\qquad &\qquad20.8\qquad &\qquad34.46\qquad &\qquad4\qquad&\qquad$4.09\times10^{21}$\qquad &\qquad$2\times10^{14}$\qquad &\qquad23.6 \qquad\\ \hline
 Case F &\qquad 53.35\qquad &\qquad22.6\qquad &\qquad30.75\qquad &\qquad11\qquad&\qquad$2.20\times10^{19}$\qquad &\qquad$1\times10^{12}$\qquad &\qquad22.3 \qquad\\ \hline
\end{tabular}
\label{tab3reh}
\end{table}
Respecting the canonical reheating ($\omega_{\text{reh}}=0$), duration of the observable inflation is related to the post-inflationary expanse \cite{Liddle:2003,Martin:2010}  through
\begin{equation}\label{deltaNfinal}
\Delta N\simeq 57.4882+\frac{1}{4}\ln\left( \frac{\varepsilon_{*}V_{*}}{\rho_{\text{end}}} \right)-\frac{1}{4}\tilde{N}_{\text{reh}}.
\end{equation}
As we mentioned heretofore, $\Delta N=N_{*}-N_{\text{end}}$, which $N_{\text{end}}$ represents  the $e$-fold number at the end of inflation, moreover  $N_{*}$,  $\varepsilon_{*}$, and  $V_{*}$ denote $e$-fold number, the value of the first slow-roll parameter, and the potential value  when the pivot scale $k_{*}=0.05{\rm Mpc}^{-1}$ leaves the horizon, respectively.

Furthermore, regarding the relation between $\rho_{\text{reh}}$  and $T_{\text{reh}}$ (reheating temperature) in the form of   $\rho_{\text{reh}}=(\pi^{2}/30)g_{*}T_{\text{reh}}^{4}$,  and applying Eq. (\ref{nktilde}) we can evaluate $T_{\text{reh}}$ as
\begin{equation}\label{Treh}
T_{\text{reh}}=\left(\frac{30}{\pi^{2}g_{*}}\rho_{\text{end}}\right)^{1/4}e^{-3\tilde{N}_{\text{reh}}/4},
\end{equation}
where $g_{*}$ designates the effective number of relativistic species upon thermalization, which is determined as   $g_{*}=106.75$   in the Standard Model of particle physics at high temperature.
It is worth noting that, reheating stage has lied  between the end of inflation and BBN, which means the constraint $10^{-2}\rm GeV\lesssim T_{\rm reh}\lesssim10^{16}\rm GeV$ can be accepted.

\begin{figure*}
\begin{minipage}[b]{1\textwidth}
\subfigure{\includegraphics[width=.48\textwidth]%
{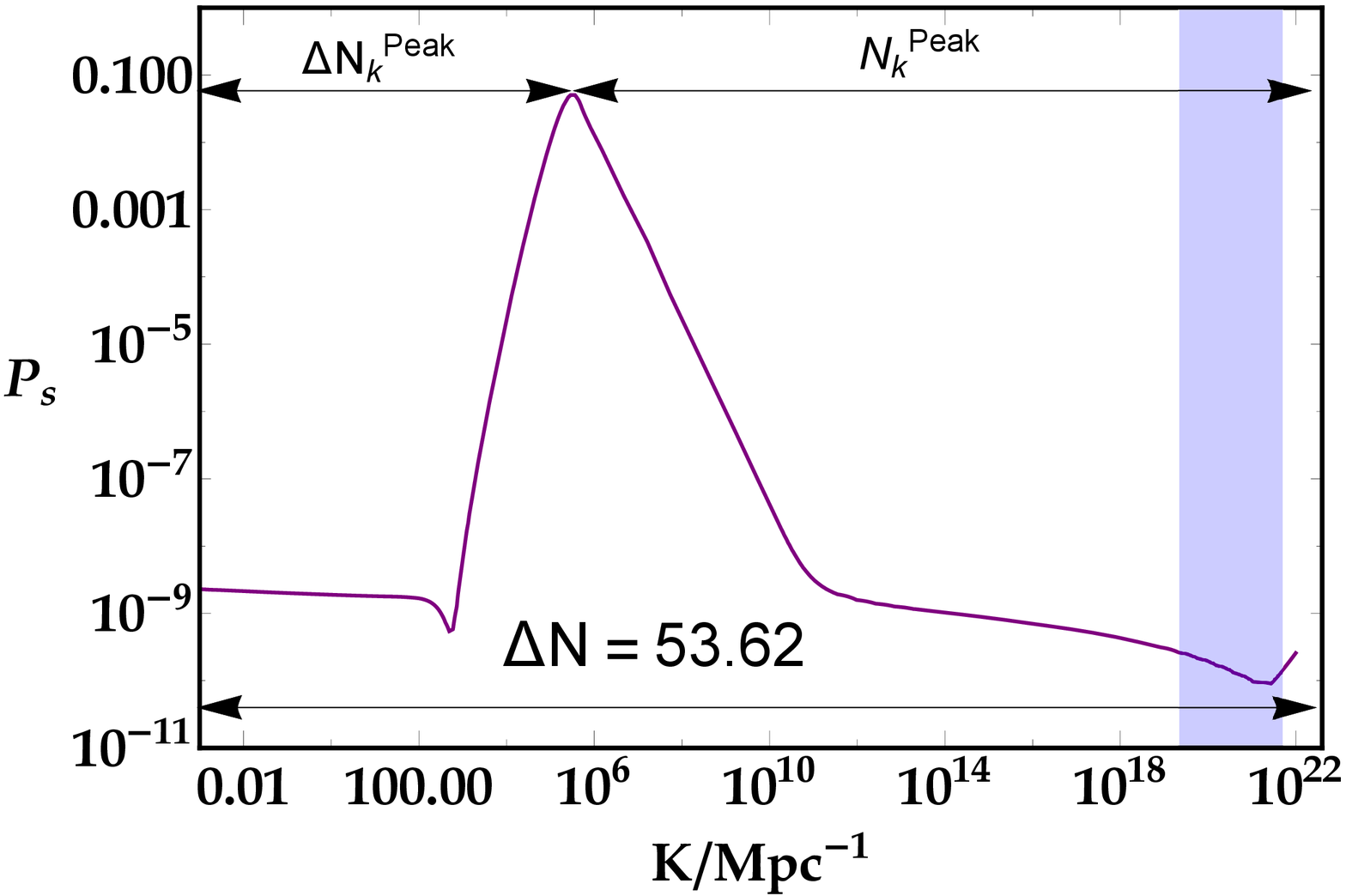}} \hspace{.1cm}
\subfigure{ \includegraphics[width=.48\textwidth]%
{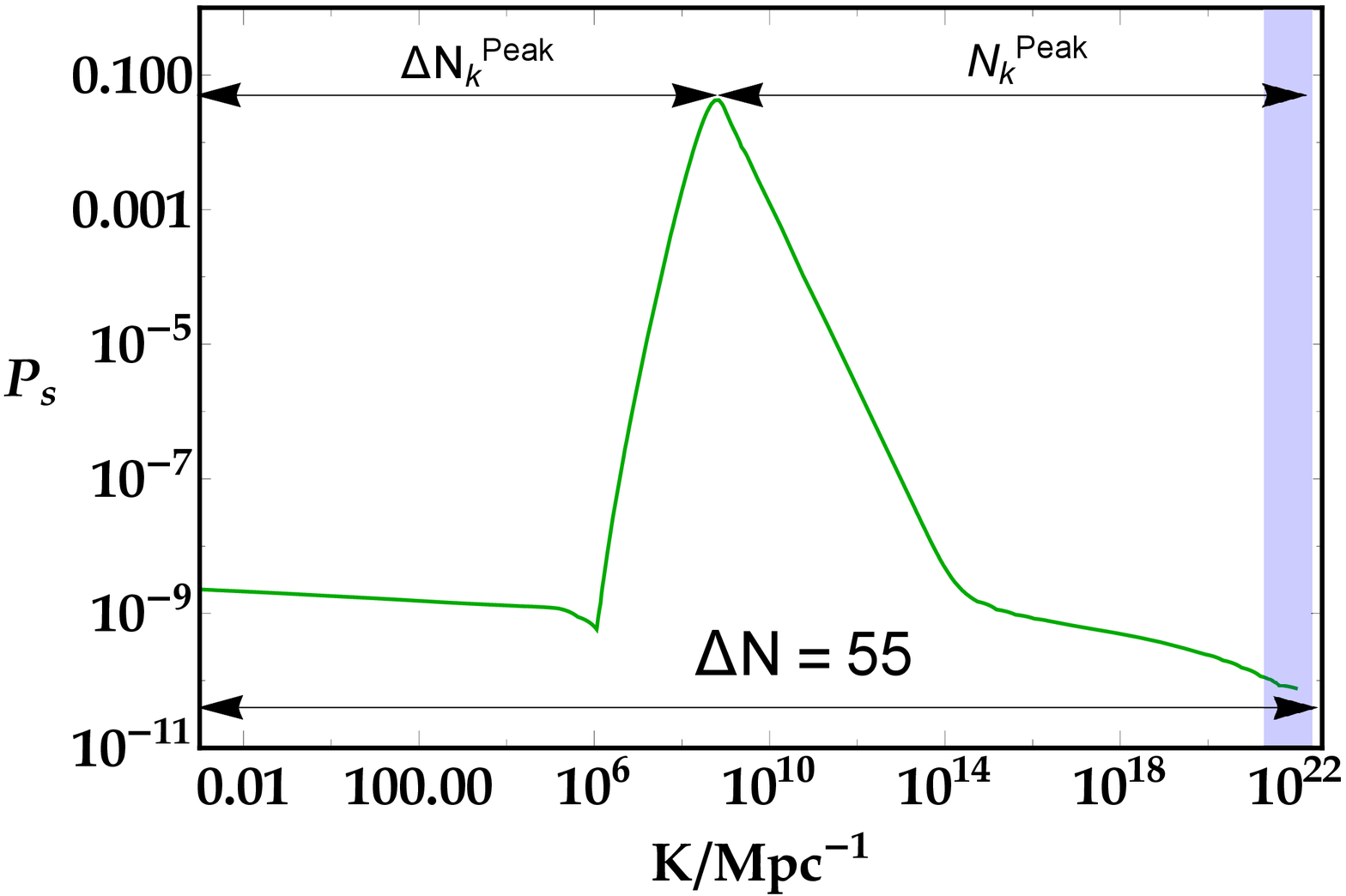}}\hspace{.1cm}
\subfigure{ \includegraphics[width=.48\textwidth]%
{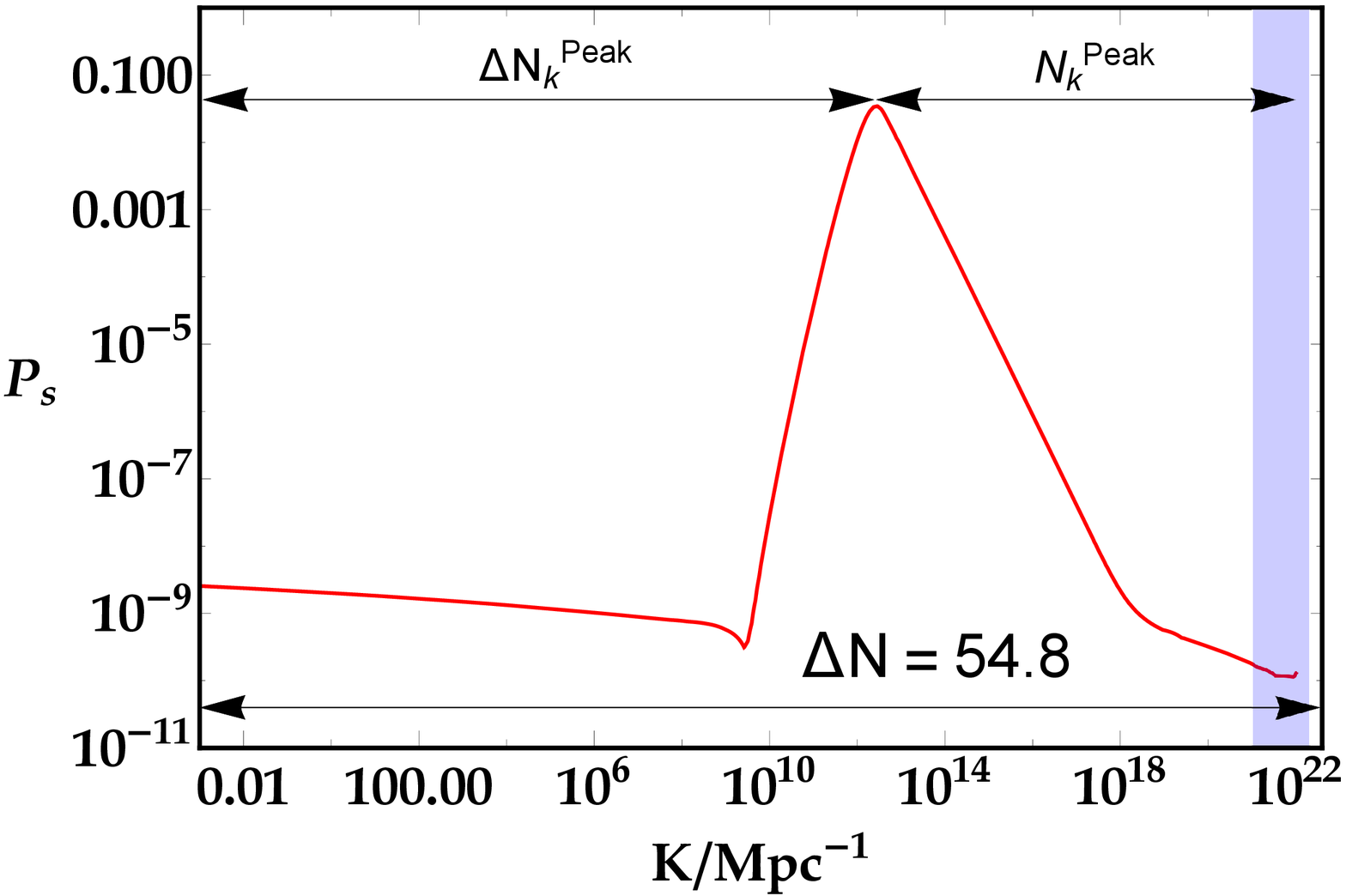}}\hspace{.1cm}
\subfigure{ \includegraphics[width=.48\textwidth]%
{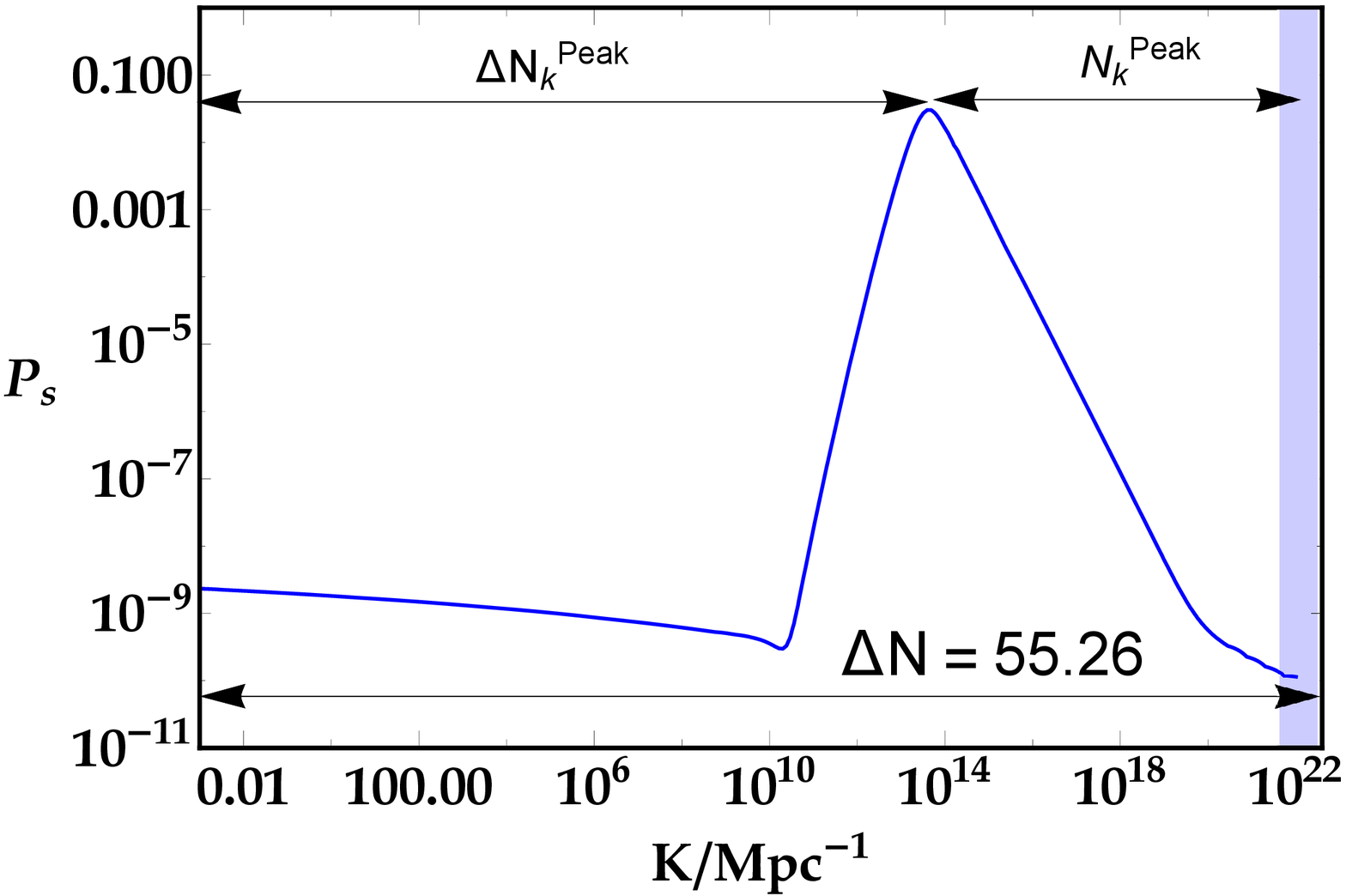}}
\subfigure{ \includegraphics[width=.48\textwidth]%
{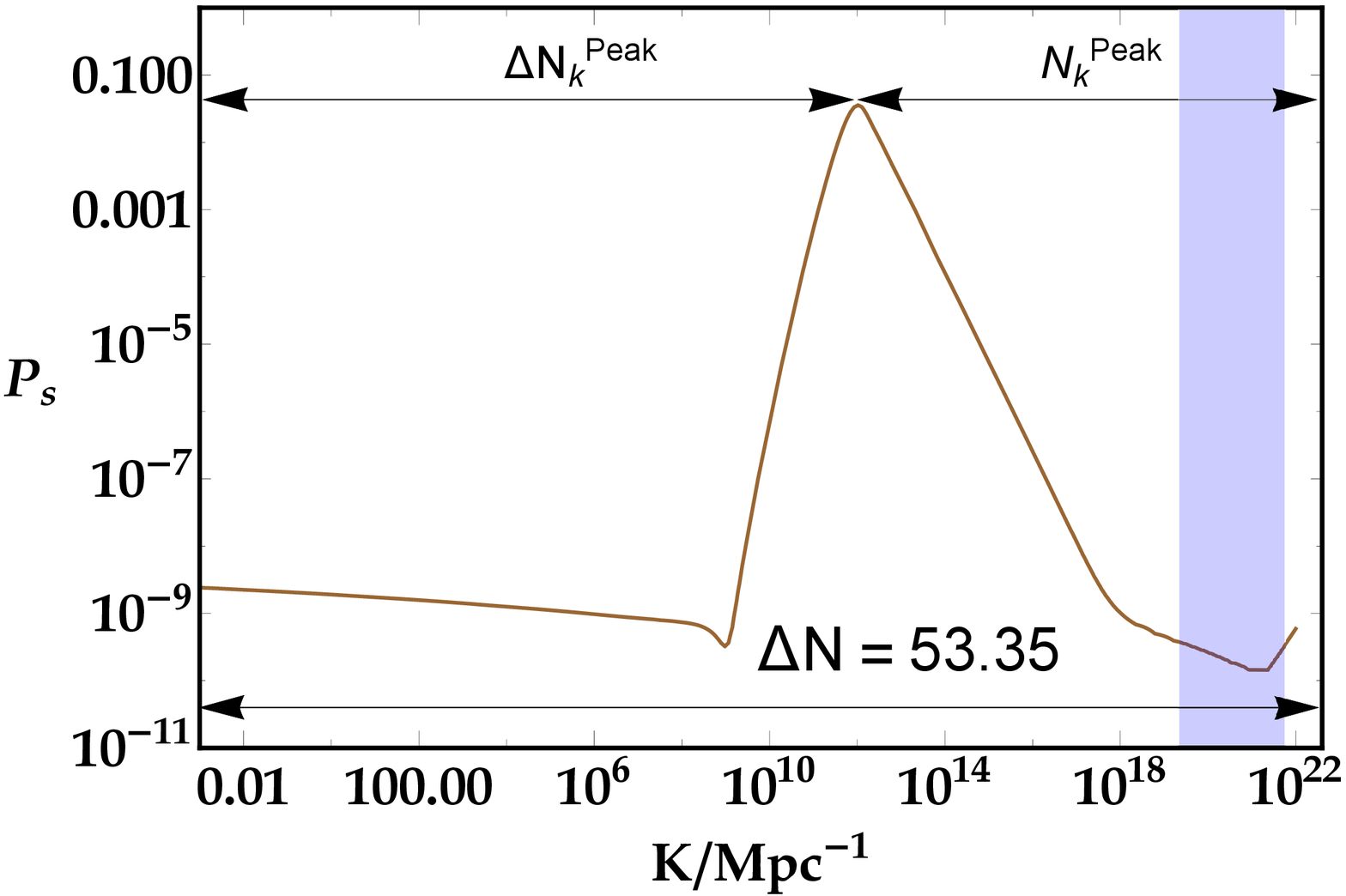}}
\end{minipage}
\caption{Evolution of the scalar power spectrum as a function of  $k$ for Case A (purple line), Case B (green line),  Case C (red line),  Case D (blue line), and Case F (brown line). The shadowy zones represent
the re-entered scales to  the horizon during the reheating stage.}
\label{fig:reh}
\end{figure*}

Now we are able to characterize  a criterion for specifying  that   the PBHs formation occurs during reheating or RD era. To this aim, in Fig. \ref{fig:reh} the duration of the observable inflation ($\Delta N$) is schemed for all Cases of Table \ref{tab1} as $\Delta N=N_{k}^{\rm peak}+\Delta N_{k}^{\rm peak}$. Note that, $\Delta N$ consists of the CMB anisotropies on large scales and scalar perturbation modes on small scales related to the PBHs generation simultaneously.

Applying Eq. (\ref{deltaNtoN}) for the scale $k_{\rm peak}^{-1}$ and taking $\omega=0$, we can write $\Delta N_{k}^{\rm peak}=\tilde{N}_{k}^{\rm peak}/2$, wherein $\tilde{N}_{k}^{\rm peak}$ denotes the post-inflationary  elapsed $e$-folds number between the end of inflation and the horizon re-entry of the scale $k_{\rm peak}^{-1}$. Suppose that the mentioned scale associated with the peak of the power spectrum re-enters the horizon upon the end of reheating stage, thus we will have $\Delta N_{\rm peak}^{(\rm cr)}=\tilde{N}_{\rm reh}/2$. Herein  duration of  the observable inflationary epoch can be written as  $\Delta N=N_{k}^{\rm peak}+ \tilde{N}_{\rm reh}/2$, then by substituting that into Eq. (\ref{deltaNfinal}) the critical value for $\Delta N_{k}^{\rm peak}$ is evaluated as follows
\begin{equation}\label{deltaNc}
\Delta N_{\rm{peak}}^{(\text{cr})}=\frac{2}{3}\left[ 57.4882+\frac{1}{4}\ln\left( \frac{\varepsilon_{*}V_{*}}{\rho_{\text{end}}} \right)-N_{k}^{\rm{peak}}\right].
\end{equation}

Our conclusions for the  values of  $\Delta N$, $N_{k}^{\rm peak}$, $\Delta N_{k}^{\rm peak}$, $\tilde{N}_{\rm reh}$, $k_{\rm reh}$, $T_{\text{reh}}$, and $\Delta N_{\rm peak}^{(\text{cr})}$  are illustrated in Table  \ref{tab3reh}  for  defined Cases in Table \ref{tab1}. Here,  $k_{\text{reh}}^{-1}$ specified as  $k_{\text{reh}}=e^{-\tilde{N}_{\text{reh}}/2}k_{\text{end}}$ \cite{Dalianis:2019}, and $k_{\text{end}}^{-1}$ are the scales  correspond to the horizon re-entry at the end of reheating and end of inflation, respectively.  Using of these results we can establish the following critical conditions for PBHs formation during RD era
\begin{align}\label{rehcondition}
&\Delta N>\Delta N^{(\text{cr})}=N_{k}^{\rm peak}+\Delta N_{\rm peak}^{(\text{cr})},\nonumber\\&
\Delta N_{k}^{\rm peak}>\Delta N_{\rm peak}^{(\text{cr})},\nonumber\\&
T_{\text{reh}}>T_{\text{reh}}^{(\text{cr})},\nonumber\\&
\tilde{N}_{\text{reh}}<\tilde{N}_{\text{reh}}^{(\text{cr})}=2\Delta N_{\rm peak}^{\text{(cr)}}.
\end{align}
 In other words, the climax scales of the scalar power spectrum re-enter the horizon and collapse to form PBHs during RD epoch provided that the above conditions are satisfied, otherwise their re-entry occurs during reheating stage.
Making a comparison between epitomized results of Table \ref{tab3reh} with their critical counterparts shows that in our study the conditions (\ref{rehcondition}) are respected, and  the scale $k_{\text{peak}}^{-1}$ comes back to the horizon and collapses to generate PBH during RD epoch.
In Fig. \ref{fig:reh} the behavior of the exact scalar power spectra in terms of $k$ for all Cases of Table \ref{tab1} is schemed. The shadowy zones symbolize the re-entered scale to the horizon during reheating stage, and it is obvious from this figure that, the climax scales comes back to the horizon during RD
epoch. Subsequently, applying the corresponding formulations with RD epoch to evaluate  mass fraction of the PBHs and energy density of the GWs is permitted in the succeeding sections.
\section{Primordial black holes Generation}\label{sec5}
\begin{figure}[H]
\centering
\includegraphics[scale=0.68]{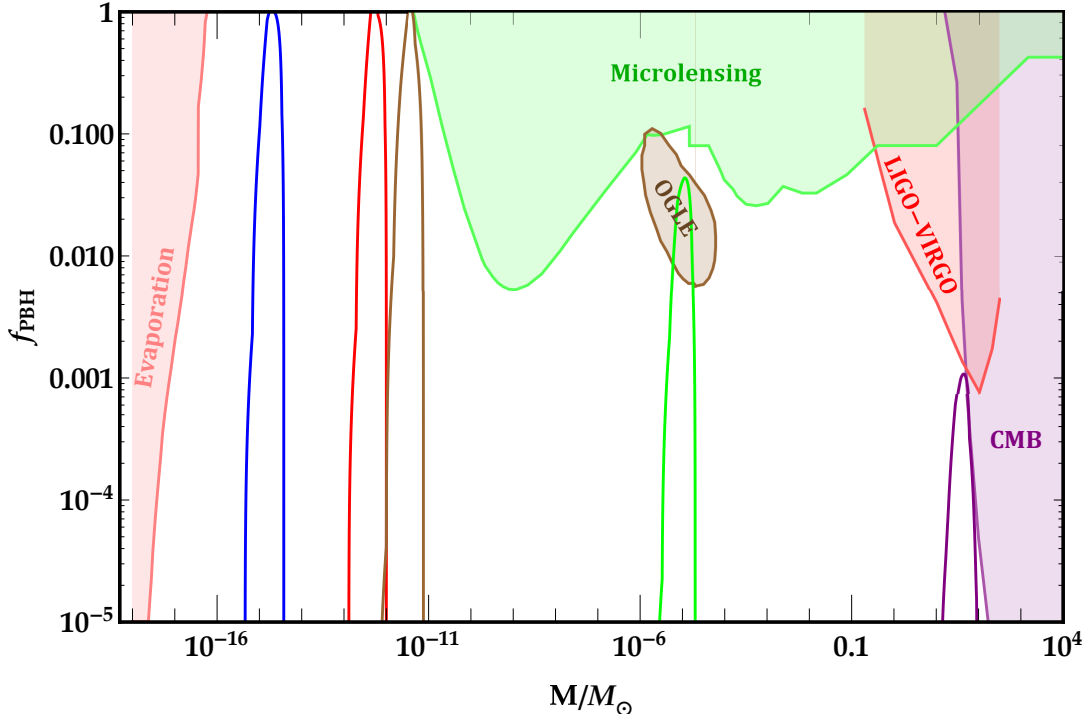}
\caption{The PBHs abundance ($f_{\rm PBH}$) in terms of PBHs mass ($M$) for Case A (purple  line), Case B (green line), Case C (red line),  Case D (blue line), and Case F (brown line). The purple area depicts the restriction on CMB from signature of spherical accretion of PBHs inside  halos  \cite{CMB}.
The border of the red shaded domain depicts the upper bound  on the PBH abundance ensued from the LIGO-VIRGO event consolidation rate \cite{Abbott:2019,Chen:2020,Boehm:2021,Kavanagh:2018}. The brown shadowy domain portrays the authorized region for PBH abundance owing to  the ultrashort-timescale microlensing events in the OGLE data \cite{OGLE}. The green shaded area allots to constraints of  microlensing events from  cooperation between MACHO \cite{MACHO}, EROS \cite{EORS}, Kepler \cite{Kepler}, Icarus \cite{Icarus}, OGLE \cite{OGLE},  and Subaru-HSC \cite{subaro}. The pink shadowy region delineates the constraints related to PBHs evaporation such as
extragalactic $\gamma$-ray background \cite{EGG},  galactic center 511 keV $\gamma$-ray line (INTEGRAL) \citep{Laha:2019}, and effects on CMB spectrum \cite{Clark}.}
\label{fig-fpbh}
\end{figure}

As we described heretofore, it is known that an enough  increase in the amplitude of curvature perturbations on
scales smaller than the CMB scales can lead to   detectable production of PBHs and GWs.
In this section, we will inquire PBHs generation in NMDC model. Pursuant to the discussions in the previous section, the re-entry of super-horizon  perturbation modes generated in the inflation epoch, occurs during RD era after inflation. If the power spectrum of these  modes have large enough amplitude, the gravity of the related overdense zones can overpower the radiation pressure and collapse to generate PBHs. The mass of generated PBH is associated with the horizon mass at the moment of re-entry through
\begin{align}\label{Mpbheq}
M_{\rm PBH}(k)=\gamma\frac{4\pi}{H}\Big|_{c_{s}k=aH} \simeq M_{\odot} \left(\frac{\gamma}{0.2} \right) \left(\frac{10.75}{g_{*}} \right)^{\frac{1}{6}} \left(\frac{k}{1.9\times 10^{6}\rm Mpc^{-1}} \right)^{-2},
\end{align}
wherein $\gamma$ is the efficiency of collapse, and in this paper we contemplate $\gamma=(\frac{1}{\sqrt{3}})^{3}$ \cite{carr:1975}. As be mentioned in the previous section  $g_{*}=106.75$ for PBHs generation in RD epoch. On the presumption that the distribution function of  perturbations abide by Gaussian statistics, and applying  Press-Schechter theory, the generation rate for PBHs with mass $M(k)$ is evaluated \cite{Tada:2019,young:2014}  as follows
\begin{equation}\label{betta}
  \beta(M)=\int_{\delta_{c}}\frac{{\rm d}\delta}{\sqrt{2\pi\sigma^{2}(M)}}e^{-\frac{\delta^{2}}{2\sigma^{2}(M)}}=\frac{1}{2}~ {\rm erfc}\left(\frac{\delta_{c}}{\sqrt{2\sigma^{2}(M)}}\right),
\end{equation}
in which "erfc" designate the error function complementary, and $\delta_{c}$ symbolizes the threshold value of the density perturbations for  PBHs generation. We contemplate $\delta_{c}=0.4$ in the way that is  recommended by \cite{Musco:2013,Harada:2013}. In addition $\sigma^{2}(M)$ denotes the coarse-grained density contrast with the smoothing scale $k$, that is specified as
\begin{equation}\label{sigma}
\sigma_{k}^{2}=\left(\frac{4}{9} \right)^{2} \int \frac{{\rm d}q}{q} W^{2}(q/k)(q/k)^{4} {\cal P}_{s}(q),
\end{equation}
wherein ${\cal P}_{s}$ is the power spectrum of curvature perturbations, and  $W$  denotes the window function which is specified as  Gaussian window $W(x)=\exp{\left(-x^{2}/2 \right)}$ in this paper.
In the light of characterizing the  abundance of PBHs,
the present fractional density parameter  of  PBHs  $(\Omega_{\rm {PBH}})$ to Dark Matter ($\Omega_{\rm{DM}})$ is evaluated as
\begin{equation}\label{fPBH}
f_{\rm{PBH}}(M)\simeq \frac{\Omega_{\rm {PBH}}}{\Omega_{\rm{DM}}}= \frac{\beta(M)}{1.84\times10^{-8}}\left(\frac{\gamma}{0.2}\right)^{3/2}\left(\frac{g_*}{10.75}\right)^{-1/4}
\left(\frac{0.12}{\Omega_{\rm{DM}}h^2}\right)
\left(\frac{M}{M_{\odot}}\right)^{-1/2},
\end{equation}
where $\Omega_{\rm {DM}}h^2\simeq0.12$ is given by Planck 2018 data \cite{akrami:2018} for current density parameter of Dark Matter.
At last we are able to calculate the PBHs abundance for all Cases of Table \ref{tab1}, using the exact power spectrum attained via resolving the Mukhanov-Sasaki equation and  Eqs. (\ref{Mpbheq})-(\ref{fPBH}). The obtained numerical and schemed results are represented in Table \ref{tab2} and Fig. \ref{fig-fpbh}. It is revealed from our conclusion that, for Case A the climax of anticipated   PBHs abundance is  located at $28.23M_{\odot}$ with $f_{\rm PBH}^{\rm peak}\simeq0.0010$, this sort of stellar-mass PBHs can
\noindent
gratify the restriction from the upper limit on the LIGO merger rate, and they are suitable candidate to expound the GWs and the LIGO events. For Case B the PBHs mass spectrum has a peak at  $M_{\rm PBH}^{\rm peak}=4.97\times10^{-6}M_{\odot}$ and $f_{\rm PBH}^{\rm peak}=0.0434$ which is placed in the permitted area by the ultrashort-timescale microlensing events in OGLE data, thus such a Case of  PBHs with earth-mass can be useful to explain microlensing events. In Cases C, D, and F our model has foretold three  PBHs abundance in asteroid-mass scales,  $M_{\rm PBH}^{\rm peak}=2.88\times10^{-13}M_{\odot}$, $M_{\rm PBH}^{\rm peak}=1.52\times10^{-15}M_{\odot}$, and $M_{\rm PBH}^{\rm peak}=2.21\times10^{-12}M_{\odot}$, with $f_{\rm PBH}^{\rm peak}$ s around $0.9832$, $0.9911$, and $1$, respectively. Hence, these Cases can be interesting candidates for composing  all  DM content of the universe.

It is worth noting that, recently Picker in \cite{Picker} has demonstrated that cosmological PBHs foretold by Thakurta metric solution in asteroid mass range of $(10^{-17}-10^{-12})M_{\odot}$ can be hotter than common  Schwarzschild case and having completely evaporated by today. On the other hand, it has been shown in \cite{Hutsi,Harada} that, the cosmological black holes and their mass growth cannot be described by Thakurta metric in the early universe. The authors of \cite{Hutsi} showed that
radiation, PBHs or particle dark matter cannot generate the requisite energy flux
for providing the mass growth of Thakurta black holes. Moreover they verified that this solution is
not applicable for binary black hole. The authors of \cite{Harada} proved that the depicted space-time by Thakurta metric has not black hole event or entrapping horizon, ergo the null energy proviso as a solution of the Einstein equation is violated. Nevertheless, this issue is currently under dispute in literatures, as shown quite recently by Kobakhidze and Picker in \cite{Kobakhidze} that, the Thakurta metric can indeed describe a cosmological black hole with apparent horizon.
\section{Induced Gravitational Waves in NMDC model}\label{sec6}
In this section, we want to study the production of  induced GWs in our model. This occurs due to re-entry  the large enough  densities of primordial perturbations to the horizon concurrent with the PBHs formation during the RD epoch. Considering the second order effects in perturbation theory, it is proven that the dynamics of tensor perturbations is originated from first order scalar perturbations. Choosing  the conformal Newtonian gauge, the perturbed FRW metric takes the form \cite{Ananda:2007}
\begin{eqnarray}
ds^2=a(\eta)^2\left\{ -(1+2\Psi)d\eta^2 +\left[(1-2\Psi)\delta_{ij}+\frac{h_{ij}}{2} \right]dx^idx^j \right\}\;,
\end{eqnarray}
where $\eta$, $\Psi$, and $h_{ij}$ indicate the conformal time, the first-order scalar perturbations, and the perturbation of the second-order transverse-traceless tensor in turn by turn. As we explained in Section \ref{sec4}, during the reheating stage after the inflationary phase the inflaton field almost completely  decay into light particles to thermalize the universe to initiate the RD era of HBB. In this regard,  the inflaton field  has a trivial influence on the cosmic evolution during RD era, and can be ignored. Accordingly, one can easily apply  the standard Einstein equation to inquire the generation of the induced GWs in RD epoch. Subsequently, the second-order tensor perturbations $h_{ij}$ gratify the following equation of motion  \cite{Ananda:2007,Baumann:2007}
\begin{eqnarray}\label{EOM_GW}
h_{ij}^{\prime\prime}+2\mathcal{H}h_{ij}^\prime - \nabla^2 h_{ij}=-4\mathcal{T}^{lm}_{ij}S_{lm}\;,
\end{eqnarray}
where  $\mathcal{H}\equiv a^{\prime}/a$  indicates the conformal Hubble parameter, and  $\mathcal{T}^{lm}_{ij}$ is  the transverse-traceless projection operator. The GW source term $S_{ij}$ is considered as
\begin{eqnarray}
S_{ij}=4\Psi\partial_i\partial_j\Psi+2\partial_i\Psi\partial_j\Psi-\frac{1}{\mathcal{H}^2}\partial_i(\mathcal{H}\Psi+\Psi^\prime)\partial_j(\mathcal{H}\Psi+\Psi^\prime)\; .
\end{eqnarray}
The scalar metric perturbation $\Psi$  in the Fourier space  during  the RD is given by  \cite{Baumann:2007}
\begin{eqnarray}
\Psi_k(\eta)=\psi_k\frac{9}{(k\eta)^2}\left(\frac{\sin(k\eta/\sqrt{3})}{k\eta/\sqrt{3}}-\cos(k\eta/\sqrt{3}) \right)\;,
\end{eqnarray}
where $k$  is the comoving wavenumber, and the primordial perturbation $\psi_k$ is associated with  the power spectrum of the curvature perturbation through the two-point correlation function as follows
\begin{eqnarray}
\langle \psi_{\bf k}\psi_{ \tilde{\bf k}}  \rangle = \frac{2\pi^2}{k^3}\left(\frac{4}{9}\mathcal{P}_{s}(k)\right)\delta(\bf{k}+ \tilde{\bf k})\;.
\end{eqnarray}
At last, the energy density of induced GWs during the RD epoch can be obtained as  \cite{Kohri:2018}
\begin{eqnarray}\label{OGW}
&\Omega_{\rm{GW}}(\eta_c,k) = \frac{1}{12} {\displaystyle \int^\infty_0 dv \int^{|1+v|}_{|1-v|}du } \left( \frac{4v^2-(1+v^2-u^2)^2}{4uv}\right)^2\mathcal{P}_{s}(ku)\mathcal{P}_{s}(kv)\left( \frac{3}{4u^3v^3}\right)^2 (u^2+v^2-3)^2\nonumber\\
&\times \left\{\left[-4uv+(u^2+v^2-3) \ln\left| \frac{3-(u+v)^2}{3-(u-v)^2}\right| \right]^2  + \pi^2(u^2+v^2-3)^2\Theta(v+u-\sqrt{3})\right\}\;,
\end{eqnarray}
wherein $\Theta$ indicates the Heaviside theta function, and $\eta_{c}$ denotes the time when the increasing of  $\Omega_{\rm{GW}}$ is halted.
The current energy spectra of the induced GWs  is associated with the  energy spectra at  $\eta_{c}$ through the following equation \cite{Inomata:2019-a}
\begin{eqnarray}\label{OGW0}
\Omega_{\rm GW_0}h^2 = 0.83\left( \frac{g_{*}}{10.75} \right)^{-1/3}\Omega_{\rm r_0}h^2\Omega_{\rm{GW}}(\eta_c,k)\;,
\end{eqnarray}
wherein $\Omega_{\rm r_0}h^2\simeq 4.2\times 10^{-5}$ is the present radiation density parameter, and $g_{*}\simeq106.75$ indicates the effective degrees of freedom in the energy density at $\eta_c$. The association between  present frequency and wavenumber can be evaluated as
\begin{eqnarray}\label{k_to_f}
f=1.546 \times 10^{-15}\left( \frac{k}{{\rm Mpc}^{-1}}\right){\rm Hz}.
\end{eqnarray}
Now we are in a position to attain the current energy  spectra of the scalar induced GWs related to PBHs, applying the actual scalar power spectrum owing to solving the MS equation, and Eqs. (\ref{OGW})-(\ref{k_to_f}) for all Cases of Table \ref{tab1}. The schemed outcomes are represented
\begin{figure}[H]
\centering
\includegraphics[scale=0.7]{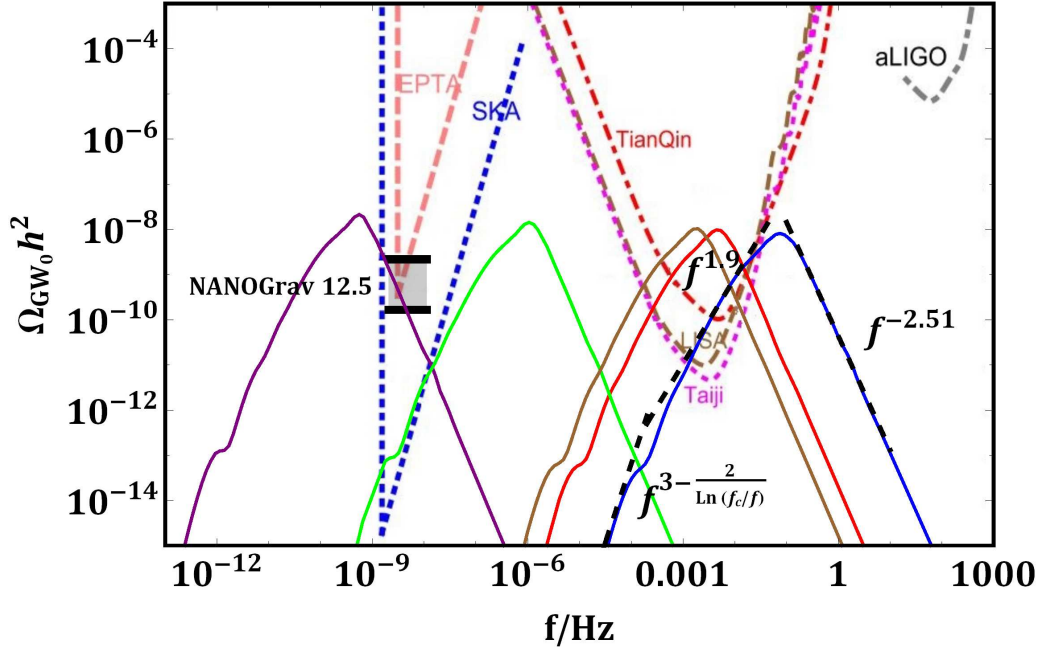}
\vspace{-0.5em}
\caption{ The present induced GWs energy density parameter ($\Omega_{\rm GW_0}$) with respect to frequency. The solid purple, green, red,  blue, and brown lines associate with Cases A, B, C,  D, and F of Table \ref{tab1},  respectively.  The  power-law form of $\Omega_{\rm GW_0}$ is illustrated  by black dashed line for Case D.}
\label{fig-omega}
\end{figure}
\noindent
in Fig. \ref{fig-omega}. Furthermore, to inquire the validity of  anticipated outcomes of our model, the sensitiveness  curves of GWs detectors comprising  European PTA (EPTA) \cite{EPTA-a,EPTA-b,EPTA-c,EPTA-d}, the Square Kilometer Array (SKA)  \cite{ska}, Advanced Laser Interferometer Gravitational Wave Observatory (aLIGO) \cite{ligo-a,ligo-b}, Laser Interferometer Space Antenna (LISA)  \cite{lisa,lisa-a}, TaiJi \cite{taiji}, and TianQin  \cite{tianqin}, are designated in the figure. Additionally $95\%$ CL from the NANOGrav 12.5 yr experiment is delineated by the  black shadowy area \cite{Nanogarv1,Nanogarv2,Nanogarv3}.
As we can see, the current density parameter spectra of induced GWs ($\Omega_{\rm GW0}$) foretold by our model,  for Cases A (purple line), B (green line), C (red line),  D (blue line), and F (brown line)   have climaxes at contrasting frequencies  with nearly identical  heights  of order $10^{-8}$. For Case A associated with stellar mass PBHs and Case B corresponding to earth mass PBHs the climaxes  of $\Omega_{\rm GW_0}$ have placed  at the frequencies  $f\sim10^{-10}\text{Hz}$ and $f\sim10^{-7}\text{Hz}$ respectively, and both Cases can be traced  via the SKA detector. Furthermore the current energy density of GWs pertinent to Case A depicted by purple line lies within the $2\sigma$ region of the NANOGrav $12.5$ yr data-set.
Moreover, the spectra of  $\Omega_{\rm GW_0}$ for  Cases C, F, and  D related to asteroid mass PBHs have climaxes localized at mHz and cHz bands which are tracked down by LISA, TaiJi, and TianQin. Inasmuch as our prognostications for all mentioned Cases could touch the sensitivity curve of GWs detectors, hence the  authenticity of our model can be assessed via the extricated data of these detectors.

Ultimately, after  inspecting the scaling of $\Omega_{\rm GW_0}$, we perceive that in the proximity of climaxes, the current density parameter spectra of induced GWs behave as a power-law function with respect to frequency ($\Omega_{\rm GW_0} (f) \sim f^{n} $) \cite{fu:2020}. In Fig. \ref{fig-omega}, this parametrization is plotted fore Case D with dashed black line. For this Case, we estimate the frequency of climax as $f_{c}=0.074{\rm Hz}$ and evaluate $\Omega_{\rm GW_0} \sim f^{1.9}$ for $f<f_{c}$, and $\Omega_{\rm GW_0} \sim f^{-2.51}$ for $f>f_{c}$. Besides, for the infrared domain $f\ll f_{c}$, we obtain a log-contingent power index as $n=3-2/\ln(f_c/f)$ which is in constancy with the analytical consequences acquired in \cite{Yuan:2020,shipi:2020}.

\section{Conclusions}\label{sec7}

Here, we investigated the feasibility of PBHs generation in a single-field inflationary model, based on the framework of nonminimal field derivative coupling to the Einstein tensor pertaining to the Horndenski theory recounted by action (\ref{action}).
The nonminimal derivative coupling to gravity leads to a brief period of ultra slow-roll inflation through gravitationally enhanced friction for the scalar field. This attribute inspired us to revise a steep potential in this framework to comprehend  a viable inflation. Thereupon, we contemplated quartic potential (\ref{v}), which is completely ruled out by Planck 2018 \cite{akrami:2018} in the standard inflationary model, and tried to recover its observational results in the NMDC framework. By assigning  the coupling parameter as a two-parted function of inflaton field (\ref{t})-(\ref{tII}), and fine-tuning of the five parameter Cases (A, B, C,  D, and F) of the model (see Table \ref{tab1}),  we could acquire  an epoch of ultra slow-roll inflation on scales smaller than CMB scale  making the inflaton  slow down, sufficient to generate PBHs. In addition, compatibility of the observational  prognostications of the model with Planck 2018 data on CMB scales is another result of these selections.

 Utilizing the exact solution of the background equations, we illustrated the behavior of inflaton field ($\phi$) and slow-roll parameters ($\varepsilon$ and $\delta_{\phi} $) in Figs. \ref{fig:phi} and \ref{fig:SRp} with respect to $e$-fold number.
It is obvious from Fig. \ref{fig:SRp} that, during the USR phase the slow-roll conditions are respected by the first parameter ($\varepsilon\ll1$), but infracted by the second one  ($\left|\delta_{\phi}\right|\gtrsim1$). Accordingly, we obtained the numerical solution of the  Mukhanov-Sasaki equation to calculate the exact power spectra of ${\cal R}$ for all Cases of Table \ref{tab1}. It is clear from the exhibited  results  in Table \ref{tab2} and  Fig. \ref{fig-ps} that, the obtained actual power spectra are compatible with the Planck 2018 data on large scales, and  have climaxes of sufficient height to generate PBHs on small scales. Furthermore, according to the observational prognostications of our model for inflation era,  the values of $r$ for all Cases and $n_{s}$ for Cases  B and F  are consistent with the $68\%$  CL of the Planck  2018  TT,TE,EE+lowE+lensing+BK14+BAO data, and the values of $n_{s }$ for Case A, Case C, and Case D  gratify the $95\%$  CL \cite{akrami:2018}. Consequently, we have been able to revive the ruled out quartic potential in the standard inflationary scenario by means of the nonminimal derivative coupling framework in our investigation.

In addition, we proved that the superhorizon climax scales of the power spectra related to PBHs formation re-enter the horizon after the reheating stage during RD era (see Fig. \ref{fig:reh}). Therefore, applying the actual ${\cal P}_{s}$ and Press-Schechter formalism we could attain detectable PBHs abundance with masses of order ${\cal O}(10)M_\odot$ for Case A (stellar mass), ${\cal O}(10^{-6})M_\odot$ for Case B (earth mass), ${\cal O}(10^{-13})M_\odot$ for Case C,  ${\cal O}(10^{-15})M_\odot$ for Case D, and ${\cal O}(10^{-12})M_\odot$ for Case F (asteroid mass). Our conclusions indicate that PBHs of Case A is suitable to describe GWs  and LIGO events,  Case  B can be useful to expound  microlensing events in OGLE data, and PBHs of  Cases C, D, and F can be interesting candidates  for composing around $98.32\%$, $99.11\%$, and $100\%$  of DM content of the universe (see Table \ref{tab2} and Fig. \ref{fig-fpbh}).

In the end, we inquired generation of the induced GWs subsequent to  PBHs formation for all Cases of our model.
Our calculation of  current density parameter spectra ($\Omega_{\rm GW_0}$) indicates that, all Cases have climaxes at contrasting frequencies  with nearly identical  heights  of order $10^{-8}$. The climaxes of $\Omega_{\rm GW_0}$ for Cases A and B have placed at frequencies  $10^{-10}\text{Hz}$ and $10^{-7}\text{Hz}$, respectively,
and both Cases can be traced  via the SKA detector. Moreover, the spectra of $\Omega_{\rm GW_0}$ for  Cases C, F, and D have climaxes localized at mHz and cHz bands which are tracked down by LISA, TaiJi, and TianQin (see Fig. \ref{fig-omega}).  Hence, validity  of our model can be assessed in view of GWs via the extricated data of these detectors. Also, we demonstrated that in the vicinity of climaxes, the spectra of  $\Omega_{\rm GW_0}$ behave as a power-law function with respect to frequency ($\Omega_{\rm GW_0} (f) \sim f^{n} $). We evaluated  $\Omega_{\rm GW0} \sim f^{1.9}$ for $f<f_{c}$ and  $\Omega_{\rm GW_0} \sim f^{-2.51}$ for $f>f_{c}$. For the infrared domain $f\ll f_{c}$   the  power index has been ascertained as $n=3-2/\ln(f_c/f)$,  which is in  constancy with the analytical consequences acquired in \cite{Yuan:2020,shipi:2020}.


\end{document}